\documentclass[notitlepage, preprint]{jfm}

\usepackage{graphicx}
\usepackage{newtxtext}
\usepackage{newtxmath}
\usepackage{natbib}
\usepackage{hyperref}
\usepackage{xcolor}
\usepackage{amsmath} 
\usepackage{multirow}
\usepackage{printlen}

\hypersetup{
    colorlinks = true,
    urlcolor   = blue,
    citecolor  = black,
}

\newcommand{\RomanNumeralCaps}[1]

\usepackage[displaymath, mathlines]{lineno}

\title{Hydrodynamics in a villi-patterned channel due to pendular-wave activity}

\author{Rohan Vernekar\aff{1}\corresp{\email{rohan.vernekar@orange.fr}},
  Faisal Ahmad\aff{1,}\aff{3},  Martin Garic \aff{2},  Dácil Idaira Yánez Martín   \aff{1,}\aff{3}, 
Claude Loverdo\aff{2}, Stéphane Tanguy\aff{3}
 \and
Clément de Loubens\aff{1}\corresp{\email{clement.de-loubens@univ-grenoble-alpes.fr}} }

\affiliation{\aff{1}Univ. Grenoble Alpes, CNRS, Grenoble INP, LRP, 38000 Grenoble, France
\aff{2} Sorbonne Université, CNRS, IBPS, JJP, 75005 Paris, France
\aff{3} Univ. Grenoble Alpes, CNRS, UMR 5525, VetAgro Sup, Grenoble INP, TIMC, 38000 Grenoble, France }

\begin{document}
\maketitle

\begin{abstract}

Inspired by small intestine motility, we investigate the flow induced by a propagating pendular-wave along the walls of a channel lined with rigid, villi-like microstructures. The villi undergo harmonic axial oscillations with a phase lag relative to their neighbours, generating travelling patterns of intervillous contraction.

Using two-dimensional lattice Boltzmann simulations, we resolve the flow within the villi zone and the lumen, sampling small to moderate Womersley numbers. We uncover a mixing boundary layer (MBL) just above the villi, composed of semi-vortical structures that travel with the imposed wave. In the lumen, an axial steady flow emerges, surprisingly oriented opposite to the wave propagation direction, contrary to canonical peristaltic flows. We attribute this flow reversal to the non-reciprocal trajectories of fluid trapped between adjacent villi, and derive a geometric scaling law that captures its magnitude in the Stokes regime.

The MBL thickness is found to depend solely on the wave kinematics given by intervillous phase lag in the low-inertia limit. Above a critical threshold, oscillatory inertia induces dynamic confinement, limiting the radial extent of the MBL and leading to non-monotonic behaviour of the axial steady flux.

We further develop an effective boundary condition at the villus tips, incorporating both steady and oscillatory components across relevant spatial scales. This framework enables coarse-grained simulations of intestinal flows without resolving individual villi.

Our results shed light on the interplay between active microstructure, pendular-wave and finite inertia in biological flows, and suggests new avenues for flow control in biomimetic and microfluidic systems.

\end{abstract}

\section{Introduction}

Inspired by the motility of smooth muscle tubular  structures, such as the gastro-intestinal tract or the ureter, the peristaltic pump stands as one of the most well established bioinspired fluidic system, enabling the efficient pumping of viscous fluids while preventing contamination from downstream to upstream \citep{esser_SilentPumpersComparative_2019}. 
The basic mechanism of peristaltic pumping takes advantage of viscous dissipation within the propagating zone of contraction, generating a pressure gradient and subsequently a net flow rate in the direction of the wave. 
Seminal theoretical work by Jaffrin and Shapiro \citep{shapiro1969peristaltic,jaffrin1971peristaltic}, rooted in thin-film flow approximation, has sparked significant interest in extending this concept across various flow regimes, for both biological \citep{sinnott2017peristaltic, Amedzrovi2022, takagi2011peristaltic} and engineering applications \citep{laser2004review, pandey2023optimal}.

Beyond peristalsis, biology continues to inspire the development of fluidic systems, particularly in microfluidics. A striking example is the emergence of artificial cilia-based devices that replicate the action of active biological ciliated cells, which measure just a few tens of micrometers \citep{den2008artificial, tabata2002ciliary, khaderi_MicrofluidicPropulsionMetachronal_2011, shields2010biomimetic}. These cilia, which line the mucosa of the respiratory system, reproductive tract, and cerebral ventricles \citep{satir2007overview, marshall2008cilia, shah2009motile}, play a crucial role in clearing viscoelastic mucus \citep{button_PericiliaryBrushPromotes_2012, loiseau2020active, choudhury_RoleViscoelasticityMucociliary_2023}.
Effective transport in a preferred direction arises from the complex motion of cilia carpets \citep{ding_MixingTransportCiliary_2014, hall_MechanicsCiliumBeating_2020}. 
At the scale of an individual cilium, the beat is inherently non-reciprocal, effectively breaking the time reversibility of low Reynolds number flows. 
At larger scales, hydrodynamic interactions promote the emergence of metachronal waves travelling across densely packed cilia carpets \citep{brennen1977fluid, dauptain2008hydrodynamics, elgeti_EmergenceMetachronalWaves_2013}, causing directional fluid pumping.
Beyond their role in biological transport, cilia-inspired microfluidic technologies also offer the ability to spatially manipulate flow into distinct regions \citep{shields2010biomimetic}: within the cilia carpet, vortical flow dominates, enhancing mixing, while above the cilia, long-range unidirectional transport is facilitated by the apparent shear stress generated at the cilia tips.

\begin{figure}
  \centerline{\includegraphics[width=1\textwidth]{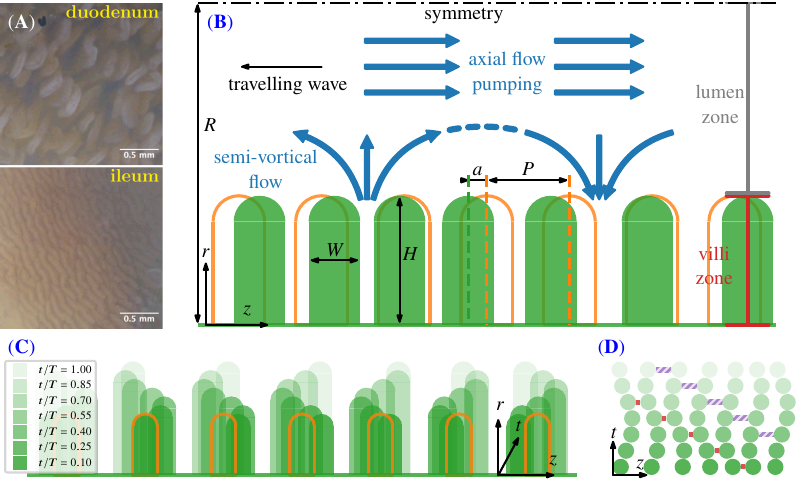}}
  \caption{\textbf{(A)}  Leaf-like and ridge-like villi of the small intestine of rat (duodenum and ileum). 
  \textbf{(B)} The planar 2D simulation domain consists of $N$ oscillating villi with periodic boundaries in the $\pm z$ direction, a wall at $r=0$, and an axis of symmetry at the channel center $r=R$. Each villus oscillates around its mean position (orange outline) with an amplitude $a$ and a phase lag $\Delta \phi$ relative to its neighbour. The mean positions of adjacent villi are separated by a constant pitch distance $P$. The instantaneous positions (filled green) of the villi show that when one section of the villi-wall contracts, the adjacent section relaxes, thereby pumping and drawing fluid in and out of the intervillous spaces. 
  \textbf{(C)} Illustration of the imposed travelling wave and time-periodic motion of the villi-wall. 
  \textbf{(D)} The same seven villi viewed from the top ($z$, $t$ axes), showing the contractions (in red) and expansions (in hatched purple) of the inter-villus gaps propagating in the $-z$ direction.}
\label{fig:schematic}
\end{figure}

While cilia-driven transport exemplifies how biological microstructures regulate fluid dynamics at small scales, another striking example of bio-inspiration is found in the small intestines.
The inner wall of the small intestine is lined with finger-like or leaf-like (or ridge-like) structures known as villi (figure \ref{fig:schematic}-A), measuring approximately $300-1000\micron$ in height \citep{garic_villi}.
Although their role in passively increasing absorption through surface area augmentation is often overstated \citep{strocchi1993role}, in-depth \textit{ex vivo} investigations reveal that these microstructures likely play an active mechanical role in enhancing mixing near the intestinal wall \citep{westergaard1986measurement,mailman1990villous,levitt1992human}.

At first glance, one might assume that densely packed intestinal villi function in a manner similar to carpets of ciliated cells in promoting fluid transport. However, from a physiological perspective, their structures are fundamentally different. A cilium is a single cellular extension whose beating pattern is actively controlled by motor proteins, whereas a villus is an elongated multicellular structure composed of an epithelial layer surrounding a network of lymphatic and vascular vessels \citep{hosoyamada2005structural}. Unlike cilia, which exhibit significant bending flexibility, villi are nearly rigid under physiological flow conditions and do not possess the same degree of freedom for bending motions \citep{lim2014determination}. Instead, their movement is driven by the contractions of the circular and longitudinal smooth muscles of the intestinal wall \citep{lentle2013mucosal}. 

Assuming that villi passively follow the dynamics of mucosal deformations induced by contractions of smooth muscle \citep{lentle2013mucosal}, one of the simplest motility patterns to consider is the forcing of villi by \textit{pendular-wave} activity. This corresponds to the propagation of longitudinal contractions along the length of the small intestine \citep{melville_LongitudinalContractionsDuodenum_1975, lammers_SpatialTemporalCoupling_2005, lentle2012comparison}. 
To model this motility, we consider villi of height $H$ as rigid microstructures \citep{lim2014determination} undergoing harmonic oscillations. The underlying \textit{pendular-wave} is manifested by an array of villi oscillating with a constant phase lag $\DPhi$ between adjacent villi (see figure \ref{fig:schematic}). The wavelength $L_z$ and the wave speed $c$ are therefore defined by the  angular frequency $\omega$ and the number of villi per wavelength, given by $N = 2\pi / \DPhi$ in this array, to enforce periodic boundary conditions.

Whereas this framework resembles the metachronal waves observed in ciliary systems dominated by viscous effects \citep{ding_MixingTransportCiliary_2014, hall_MechanicsCiliumBeating_2020}, the villi are at least an order of magnitude larger than cilia, and thus inertial effects are expected to significantly influence the flow field induced by the oscillatory activity of a villus array \citep{wang2017three, puthumana2022steady}. 
In flows driven by oscillatory boundaries, fluid inertia introduces an additional hydrodynamic length scale, the Stokes layer thickness, defined as $\StLayer = \sqrt{\nu/\omega}$, where $\nu$ is the kinematic viscosity \citep{schlichting_BoundaryLayerTheory_1960}. This length scale directly influences both the location and strength of the steady streaming flow (SSF), an effect with important implications for mixing in microfluidic systems \citep{costalonga2015low, fishman2022mixing}. 
Taking the villus width $W$ as the characteristic geometric length scale, the Womersley number defined as $\Wo = W / \StLayer$, is of order unity under physiological conditions, confirming a departure from the Stokes flow limit (see Table~\ref{tab:ratDuodenumValues}).
Although SSF is typically associated with inertial effects, it can also arise under Stokes flow conditions when additional degrees of freedom, such as boundary deformability, are present \citep{marmottant2024large, cui_ThreedimensionalSoftStreaming_2024}. Cilia-driven flows, for example, can be interpreted as a form of SSF in the Stokes regime, although this classification is not commonly adopted in the literature \citep{Riley2001Jan, hall_MechanicsCiliumBeating_2020}.

Our aim is to decipher the physical mechanisms that govern the spatio-temporal organization of the flow when villi-like microstructures are forced by \textit{pendular-wave} activity.
In Section~\ref{sec:probStatement}, we define the problem and perform the non-dimensionalization. The lattice Boltzmann method (LBM), used to compute the flow fields, is described in Section~\ref{sec:lbm}, along with a high-accuracy treatment of the moving boundaries required to capture secondary flow phenomena such as SSF. Section~\ref{sec:nonPropagating} presents the flow generated by non-propagating villus oscillations, a situation in which SSF manifests through inertial effects, and whose magnitude is several orders lower than that of the instantaneous flow. In Section~\ref{sec:propagating}, we show that a phase-lagged travelling-wave villus motion produces net directional transport and the emergence of a near-wall mixing boundary layer (MBL). Section~\ref{sec:MBL} explores the structure and evolution of the MBL over a range of Womersley numbers from small to moderate values. In Section~\ref{sec:fluxesScaling}, we explain the physical origin of the axial steady flow opposing the pendular-wave and propose a mechanistic scaling law that captures its magnitude and direction. Section~\ref{sec:BCeffective} introduces an effective boundary condition at the villus tips that reproduces the key flow features without resolving individual villi. Finally, conclusions and perspectives are presented in Section~\ref{sec:conclusions}.

\section{Problem statement and non-dimensionalization}
\label{sec:probStatement}

\subsection{Problem statement}
\label{subsec:probStatement:state}

To explore the physical mechanisms underlying the spatio-temporal organization of the flow, we performed a systematic parametric study using a symmetric 2D planar channel of diameter $2R$, whose walls are patterned with moving ridge-like villus structures of uniform height $H$ and width $W$, as illustrated in figure~\ref{fig:schematic}. Previous 3D simulations of stationary contractions by \cite{lim2015flow}, with different 3D villus shapes, show only minor qualitative differences in the resulting flow fields.

The fluid flow is solved both in the intervillous spaces and in the main channel lumen. The villi are modelled as rigid structures.
To validate this assumption, we carry out a simple order of magnitude calculation considering a single villus as a slender cantilever beam of length $L \approx 500 \micron$ and square cross-section of width $W \approx 200 \micron$.
The villus beam, made up of healthy intestinal extracellular matrix, is given to posses a Young's modulus of $\approx 2.9$kPa \citep{creff_VitroModelsIntestinal_2021}.
At a large chyme viscosity of $100$~mPa$\cdot$s, a maximum wall strain rate of $\approx 0.47$s\textsuperscript{-1} was computed for physiologically mapped pendular motion in the rat gut \citep{de2013fluid}. 
Using the slender beam theory, purely viscous stresses here would result in a maximum villus tip deflection of $< 0.5 \micron$.
This is $< 0.1 $\% of the total villus length, validating the rigidity assumption for the villi.
The rigid villi condition is further supported by two experimental studies \citep{lim2014determination, lentle2013mucosal}, which demonstrate that villus motion is driven by contractions of the underlying muscularis mucosa and does not result from bending deformation under physiological fluid shear stresses. 
Note that in the brushtail possum, smooth muscle contractions induce wrinkling of the mucosa \citep{lentle2013mucosal}. We neglect this phenomenon in the present model, as our ongoing experiments in the rat duodenum have not shown such wrinkling, suggesting that it may be species-dependent.

The spatial coordinates along the axial and radial dimensions of the channel are given by $\bm{x} = (z, r)$, respectively.
Along $r$, the channel is bounded by a moving wall at $r = 0$, detailed with $N$ periodic villi, and a symmetry boundary at $r = R$.
The domain is periodic along $\pm{z}$.
For the purpose of analysis the domain is parted into two zones; the region patterned with villi where $0\le r \le H$ is termed the \textit{villi zone}, and the channel space where $H < r \le R$, is termed the \textit{lumen zone}.
The villi harmonically oscillate along $z$, in time period $T$, around their mean axial positions as shown in figure~\ref{fig:schematic}(C). The mean villi positions (orange outlines in figure~\ref{fig:schematic}) are equispaced by distance $P$ along $\bm{z}$, therefore giving a periodic channel length of $L_z=NP$.

Each individual villus, indexed $i=1,2\dots N$, is put into axial translation by propagating  contractions of the longitudinal smooth muscle cells of the small intestine \citep{lentle2012comparison}. Consequently, the $\ith$ villi velocity is modelled as,
\begin{equation}
	\label{eq:villiOscVel}
    \bm{U}_i(t) = \omega a \sin \left( \omega t + (i-1) \Delta \phi \right) \hat{\bm{z}},
\end{equation}
and the $\ith$ villi axial position is,
\begin{equation}
	\label{eq:villiOscPos}
	\bm{X}_i(t) = \overline{\bm{X}}_i - a \cos \left( \omega t + (i-1) \Delta \phi \right) \hat{\bm{z}},
\end{equation}
\noindent
where, $\omega=2 \pi f$ is the circular frequency, $f=1/T$ is the oscillation frequency, $a$ is the oscillation amplitude, $t$ is the time, $\DPhi$ is the phase lag between adjacent villi, $\overline{\bm{X}}_i$ is the mean axial villi position and $\hat{\bm{z}}$ is the unit vector along the axial direction.

Since we model a finite number of villi, we can access only discrete values for the phase lag $\DPhi = 2 \pi /N$.
When $N = 1$, $\DPhi = 2 \pi$, meaning all villi are moving in-phase. 
This is equivalent to the case where $N$ becomes large ($N \to \infty$), and $\DPhi \to 0$, a situation recently studied in 3D by \cite{puthumana2022steady}.
When $N = 2$, $\DPhi = \pi$, and adjacent villi move with velocities exactly out-of-phase. 
In the intermediate situation, i.e.\ $0 < \DPhi < \pi$ (or $3 \le N < \infty$), a propagating wave of intervillous contraction is generated along the villi-wall.
We term this as the \textit{pendular-wave} activity.
This wave travels in the $-\hat{\bm{z}}$ direction at a speed of $c=L_z/T =  2 \pi f P / \DPhi$.
Figure~\ref{fig:schematic}(D) demonstrates how contractions and relaxations along the intervillous spaces (see marked rectangular regions) travel from right to left along $-z$.
The bottom walls of the intervillous gaps are assigned instantaneous velocities that are linearly interpolated between those imposed by their bounding villi.

The fluid is assumed Newtonian, with dynamic viscosity $\mu$ and mass density $\rho$.
The intestinal fluids are generally highly heterogeneous, and the Newtonian assumption provides a well-defined baseline in the regime of low particulate volume fraction, particularly in the proximal part of the small intestine where the digesta is mixed with a large quantity of secretions \citep{hardacre2018predicting}. This assumption, which is also relevant in view of potential microfluidic applications, facilitates the elucidation of the underlying physical mechanisms and the comparison across studied parameters.
The flow is modelled by the incompressible Navier-Stokes equations given by,
\begin{equation}
	\label{eq:NSincompress} 
	\bm{\nabla} \cdot \bm{u} = 0,
\end{equation}
\begin{equation}
	\label{eq:NSmomentum}	
	\rho \frac{\partial {\bm{u}}}{\partial {t}} + \rho{\bm{u}} \cdot \bm{\nabla} \bm{u} = - \bm{\nabla} p + \mu \nabla^2 \bm{u},
\end{equation}
\noindent
where $p$ is the pressure and $\bm{u}$ the fluid velocity vector field.

\subsection{Non-dimensionalization}
\label{subsec:probStatement:ND}
One observes that there are multiple length and time scales in the problem, as seen in  table~\ref{tab:ratDuodenumValues}. We can identify time scales for four physical processes:
(i) the villi oscillation timescale $t^{{\omega}} = 1 / \omega$, (ii) the fluid advective timescale $t^{adv} = W / (\omega a)$, (iii) the viscous timescale $t^{\mu} = W^2 \rho / \mu$, and (iv) the time scale imposed by the boundary travelling wave $t^{c} = W/c = W \DPhi / (\omega P)$.

Furthermore, oscillatory boundaries introduce a characteristic length scale for viscous dissipation normal to the wall, termed as the Stokes layer, $\StLayer = \sqrt{\mu / (\omega \rho)}$ \citep{schlichting_BoundaryLayerTheory_1960}.
A priori, it is unclear which length and time scales would dominate the problem at hand. In the above, we have chosen the villi width $W$ as the length scale of the problem, in line with the literature \citep{puthumana2022steady, tatsuno1973circulatory, kotas_visualization_2007}.
We make the choice for non-dimensionalization as: $\tilde{\bm{u}} = \bm{u}/(\omega a)$, $\tilde{p} = p W/(\mu \omega a)$, $\tilde{\bm{x}} = \bm{x}/W$ and $\tilde{t}=t \omega$ for the typical velocity, pressure, length and time, respectively, to obtain the non-dimensional momentum equation as,
\begin{equation}
\label{eq:NSmomentumND}	
\frac{1}{\tilde{a}} \frac{\partial \tilde{\bm{u}}}{\partial \tilde{t}} + \tilde{\bm{u}} \cdot \tilde{\bm{\nabla}} \tilde{\bm{u}} = - \frac{1}{\tilde{a} \Wo^2} \tilde{\bm{\nabla}} \tilde{p} + \frac{1}{\tilde{a} \Wo^2} \tilde{\nabla}^2 \tilde{\bm{u}}.
\end{equation}
\noindent
Two non-dimensional numbers emerge in equation (\ref{eq:NSmomentumND}), namely the Womersley number $\Wo=W/\StLayer$ \citep{loudon_UseDimensionlessWomersley_1998} and the reduced oscillation amplitude $\tilde{a}=a/W$.
The villi motility in (\ref{eq:villiOscVel}) can now be similarly re-written in the dimensionless form as,
\begin{equation}
	\label{eq:villiOscVelND}
	\tilde{\bm{U}}_i (\tilde{t}) = \sin \left( \tilde{t} + (i-1) \DPhi \right) \hat{\bm{z}}.
\end{equation}

Three parameters then govern the problem, namely the boundary imposed phase lag $\DPhi$, the Womersley number $\Wo = W/\StLayer$ and the reduced amplitude $\tilde{a}=a/W$. 

The geometry of villi varies both within a single species and across different species, as listed in Table~\ref{tab:ratDuodenumValues}. However, the differences observed within the same region are of the same order of magnitude. Therefore, it is not relevant to study large variations in the geometric ratios. We thus fix these ratios to the typical values reported for the rat small intestine, i.e. $R/H = 5.6$, $H/W = 2.5$, and $P/W = 1.6$, for all simulations \citep{lentle2012comparison, hosoyamada2005structural, casselbrant_AsymmetricMucosalStructure_2022}.

The longitudinal muscular activity is reported at $\sim$0.5 Hz in various species (table~\ref{tab:ratDuodenumValues}).
Based on this value, the typical physiological Stokes layer thickness $\StLayer$ is estimated at least 470~$\micron$ for a watery digesta.
Consequently, for typical rat villi lengths of $\approx 500 \micron$, the typical physiological $\Wo = W/\StLayer \approx 0.42$. 
There is no direct measurement for $\DPhi$ reported in the literature; we estimate it to be small, with an upper bound of 0.2.
In this paper we go beyond the limited physiological range of values of $\Wo$ and $\DPhi$, in order to investigate the physical mechanisms governing the flow patterns and to assess their potential for microfluidic applications.
$\Wo$ is varied over three orders of magnitude and $\DPhi$ from $0$ to $\pi$ in discrete steps. 
Regarding the amplitude of oscillations, we restrict our study to the small-amplitude regime, with $\tilde{a} = 0.1$ and $0.2$. This choice is motivated by previous studies on steady streaming flows, which have shown that more complex flow structures can emerge in the large-amplitude regime, even at low $\Wo$ \citep{tatsuno1973circulatory}. When $\Wo \ll 1$, along with $\tilde{a}\ll1$, we see that the equation (\ref{eq:NSmomentumND}) reduces to the steady Stokes equation. However, when $\Wo \gg 1$ we expect an inertial flow regime, with all terms in (\ref{eq:NSmomentumND}) playing a role.

The results presented in this paper are based on a total of 128 simulations spanning the parameter space $(\Wo,\, \DPhi,\, \tilde{a})$.

\begin{table}
  \begin{center}
\def~{\hphantom{0}}
  \begin{tabular}{lcccc}
  	Radius of the small intestine   & $R$ & 3--30 mm & \citep{lammers_SpatialTemporalCoupling_2005, lentle2012comparison, lentle2013mucosal}\\
	Width of the villi   & $W$ & 50 -- 400 $\mu$m & \citep{garic_villi} \\
	Length of the villi   & $H$ &  300 -- 1000 $\mu$m & \citep{garic_villi} \\
	Pitch   & $P$ & 30 -- 480 $\mu$m & \citep{garic_villi} \\
    Frequency   & $f$ & 0.1 -- 0.7 Hz & \citep{lammers_SpatialTemporalCoupling_2005, lentle2012comparison, lentle2013mucosal}  \\ 
    Wave speed   & $c$ & 10 -- 30 mm/s & \citep{lammers_SpatialTemporalCoupling_2005, lentle2012comparison} \\
    Velocity amplitude    & $U_0$ & 0.2 -- 5 mm/s & \citep{de2013fluid,fullard_PropagatingLongitudinalContractions_2014} \\
    Displacement amplitude    & $a=U_0/\omega$ & 50 -- 1000 $\mu$m & \citep{de2013fluid,fullard_PropagatingLongitudinalContractions_2014}\\
   
    Fluid density   & $\rho$ & $\sim$994 kg/m$^{3}$ & water at 37$^{\circ}$C \\
    Fluid viscosity    & $\mu$ &  $\ge$ 0.7 mPa.s & water at 37$^{\circ}$C\\
    Stokes layer  & $\StLayer=\sqrt{{ \mu}/{\omega \rho} }$ & 400 -- 1000 $\mu$m & estimated\\
    Womersley number & $\Wo = W /\StLayer$ & 0.05 -- 1.0 & estimated \\
    Phase lag  & $\DPhi=\omega P/c$ & 10$^{-5}$ -- 10$^{-1}$& estimated \\
  \end{tabular}
  \caption{Typical values characterizing villus geometry in the small intestine (mouse, rat, possum, rabbit, chicken, human, horse; see details in \citeauthor{garic_villi} 2025) and longitudinal motility in rat \citep{lentle2012comparison,de2013fluid}, possum \citep{lentle2013mucosal}, and rabbit \citep{lammers_SpatialTemporalCoupling_2005,fullard_PropagatingLongitudinalContractions_2014}. Orders of magnitude of dimensionless parameters are estimated based on these physiological values.}
  \label{tab:ratDuodenumValues}
  \end{center}
\end{table}

\section{Numerical methods}
\label{sec:lbm}

\subsection{Two-relaxation-time lattice Boltzmann solver}

We solve the incompressible Navier-Stokes equations using the lattice Boltzmann method (LBM) using the D2Q9 lattice, which discretizes coordinate space in two directions and the velocity space in nine directions \citep{he_TheoryLatticeBoltzmann_1997, zhang_LatticeBoltzmannMethod_2011a}. The simulation code is validated for the Stokes second problem constrained to a symmetric channel to be second-order accurate with the comparisons shown in figure S1, and is open sourced \citep{vernekar20253dintestinalflow}.
One key advantage of LBM is that it solves the flow on a fixed Cartesian grid, which remains efficient even in the presence of moving boundaries. This grid-based formulation allows for straightforward parallelization, as most operations are local.
Moreover, we use the two-relaxation-time (TRT) scheme, which significantly reduces the relaxation-time dependent error in the LBM \citep{ginzburg_TworelaxationtimeLatticeBoltzmann_2008}, in order to accurately capture second order flow phenomena. In the TRT-LBM, ``distribution function'' (or  ``population'') $g$ is updated at every lattice node according to the following,
\begin{align}
	\label{eq:postCollLBTRT}
	&g_i^{\ast}(t^{\ast}, \bm{x}) = g_i(t,\bm{x}) + \frac{1}{\tau^+} (g^{eq+}_i(t,\bm{x}) - g^+_i(t,x)) + \frac{1}{\tau^-} (g^{eq-}_i(t,\bm{x}) - g^-_i(t,\bm{x})),\\
	\label{eq:postStreamLBTRT}
	&g_i(t+\Dt, \bm{x+\Delta x}) = g_i^{\ast}(t^{\ast}, \bm{x}),
\end{align}
\noindent
where, $\bm{x}$ is the position coordinate, $t$ the time, and the index $i=0,1,\dots,8$ indicates discrete lattice velocity directions. For the $\ith$ direction, $\bm{x}+\bm{\Delta x} = \bm{x}+\bm{e}_i\Dt$, where $\bm{e}_i $ is the lattice velocity vector \citep{kruger_LatticeBoltzmannMethod_2017}. 
$\tau^+$ and $\tau^-$ are the two relaxation times associated with symmetric ($^+$) and anti-symmetric ($^-$) parts of the populations which are defined as,
\begin{equation}
	\label{eq:symmAntiSymmPop}
	g^+_i = \frac{g_i + g_{\bar{i}}}{2},
	~g^-_i = \frac{g_i - g_{\bar{i}}}{2},
	~g^{eq+}_i = \frac{g^{eq}_i + g^{eq}_{\bar{i}}}{2},
	~g^{eq-}_i = \frac{g^{eq}_i - g^{eq}_{\bar{i}}}{2},
\end{equation}
\noindent
where the direction $\bar{i}$ is defined such that $\bm{e}_{\bar{i}} = - \bm{e}_i $.

We use the \cite{he_LatticeBoltzmannModel_1997} incompressible equilibrium, which is given below,
\begin{equation}
	\label{eq:HeLuoEquilibroum}
	g^{eq}_i (\bm{x}, t) = w_i \rho +  w_i \rho_0 \left(\frac{\bm{u}\cdot \bm{e}_i}{c_s^2} + \frac{(\bm{u}\cdot \bm{e}_i)^2}{2 c_s^4} + \frac{\bm{u}\cdot\bm{u}}{2 c_s^2} \right),
\end{equation}
\noindent
where $w_0=4/9, w_{1-4}=1/9$ and $w_{5-8}=1/36$ are the lattice direction weights, and $\rho = \rho_0 + \delta \rho$ is the fluid density, taken as the sum of constant and variable parts.
For the D2Q9 lattice discretization, $c_s= \Delta x/(\Dt \sqrt{3}) $ is is the so called ``lattice sound speed''.
In simulation (or lattice) units, we take the nodal distance $\Delta x = 1$, the time step $\Dt = 1$ and $\rho_0=1$.
Using the incompressible equilibrium rather than the more popular compressible equilibrium is important to accurately capture the time-integrated steady streaming flow.

The macroscopic variables in the solution are recovered through,
\begin{align}
	\label{eq:macroVarLB}
	\rho = \sum_i g_i ~ &\mathrm{and} ~ \bm{u} = \frac{1}{\rho_0} \sum_i \bm{e}_i g_i,\\
	\label{eq:pressFromDen}
	p=c_s^2 \rho~ &\mathrm{and}~p_0=c_s^2 \rho_0,
\end{align}
\noindent
where $p$ gives the fluid pressure, and $p_0$ is the datum pressure.
Through Chapman-Enskog analysis, the kinematic viscosity is related to the relaxation time as $\nu = c_s^2 (\tau^{+} - \Dt/2)$ \citep{ginzburg_TworelaxationtimeLatticeBoltzmann_2008}.

The LBM algorithm follows two simple steps; first, ``collision'' where (\ref{eq:postCollLBTRT}) is evaluated at some intermediate time $t^{\ast} > t$, followed by the ``streaming'' equation  (\ref{eq:postStreamLBTRT}), where the post-collision populations advance to their neighbouring nodes at the end of $t+\Dt$.
This completes one time step $\Dt$ at the end of which the macroscopic variables $\bm{u}$ and $p$ are computed from (\ref{eq:macroVarLB}) and (\ref{eq:pressFromDen}).
These are then used to compute equilibrium distribution $g^{eq}$ from (\ref{eq:HeLuoEquilibroum}) for the next time step.

\subsection{Moving boundary condition}
\label{sec:lbm:sub:movingBC}
Contrary to the LB simulations involving cilia \citep{hall_MechanicsCiliumBeating_2020}, where the elongated microstructures are thinner than one lattice unit, villi have a finite thickness that spans several lattice nodes.
The complex boundary in this study demands a higher-order LB scheme for treating moving boundaries in order to obtain an accurate solution for the instantaneous and steady streaming flows around the villi and in the lumen.
It is equally important to adopt a robust ``fresh'' node treatment alongside the scheme, in order to ensure a stable solution, and prevent degradation of accuracy.
Fresh nodes are those lattice nodes that transition from inside the villi into the fluid zone as the villi boundary moves, and where $\rho$ and $\bm{u}$ need to be guessed.
We therefore use a second-order accurate interpolated bounce-back scheme (IBB), alongside an iterative procedure to re-fill values for fresh fluid nodes \citep{ginzburg_TworelaxationtimeLatticeBoltzmann_2008, chen_ComparativeStudyLattice_2014}.

The IBB scheme is adapted from the family of linear interpolation (LI) schemes that are computationally local at a lattice node \citep{ginzburg_UnifiedDirectionalParabolicaccurate_2023}.
If the boundary node on the fluid side is at $\bm{x}_F$, and the villi wall lies at $\bm{x}_W = \bm{x}_F + q \bm{e}_i$, where $q$ is the fractional distance to the wall along direction $i$, the unknown population along $\bar{i}$ to be streamed from the wall to node at $\bm{x}_F$ is computed as,
\begin{align}
	\label{eq:IBB_LI}
	g_{\bar{i}} (\bm{x}_F, t+\Dt) &= a_1 g^{\ast}_i (\bm{x}_F, t^{\ast}) + a_2 g^{\ast}_{\bar{i}} (\bm{x}_F, t^{\ast}) + \mathcal{A}_i + \mathcal{W}_i, \\
	\mathcal{A}_i &=  a_3 \left( I^{FF} g_i (\bm{x}_F, t+\Dt) + (1-I^{FF}) g_i (\bm{x}_F, t) \right), \\ 
	\mathcal{W}_i &= a_0 \frac{w_i}{c_s^2} \rho_0 \frac{\bm{u}_W(t) + \bm{u}_W(t+\Dt)}{2} \cdot \bm{e}_i,
\end{align}
\noindent
where $a_{0-3}$ are interpolation coefficients and $\bm{u}_W$ is the wall velocity.
We set $I^{FF} = 1$ when $\bm{x}_{FF} = \bm{x}_{F} - \bm{e}_i \Dt$ is a fluid node, whereas $I^{FF} = 0$ when $\bm{x}_{FF}$ is a solid node (i.e.\ lies within the villi).

The interpolation weights are set according to the following,
\begin{equation}
	\label{eq:IBB-LI_interpWeights}
	a_0 = \frac{3}{5} \left( \frac{4}{1+2q} \right), ~a_1 = a_0 \left( \frac{1}{2} + q \right) -1, ~a_2 = 1 - \frac{1}{2} a_0, ~a_3 = 1 - (a_1 + a_2).
\end{equation}
\noindent
Note that in (\ref{eq:IBB-LI_interpWeights}), the prefactor for $a_0$ is set to $(3/5)$ for reasons of simulation stability and to minimising velocity oscillations due to fresh node transitions (though this prefactor value can be varied between $[0, 1]$).
For all reported simulations we take 20 lattice nodes for the width $W$ of each villi and 32 lattice nodes for the pitch $P$.

\subsection{Fresh node treatment}
\label{sec:lbm:sub:movingBC:freshNodeTreat}

The process of re-initializing all $g_i$ values at the nodes that are uncovered into the fluid side (as
the villi move) is called as fresh node treatment.
This is a major source of error as well as of numerical velocity oscillations in our computations, and therefore needs to be handled in a robust manner \citep{chen_ComparativeStudyLattice_2014,ginzburg_LatticeBoltzmannMethod_2025}.
We adopt the local iteration refill (LIR) procedure, with slight modifications, which is highly effective at countering these shortcomings \citep{tao_InvestigationMomentumExchange_2016}.
In the LIR we first identify groups of connected fresh nodes (at least one fresh node neighbour along $\bm{e}_i$), and carry out the following procedure:
\begin{enumerate}
	\item Execute local collision step for all link-wise fluid neighbour nodes identified for a fresh node group using (\ref{eq:postCollLBTRT}).
	\item Fresh nodes are partially re-filled by streaming-in post-collision populations from their neighbour nodes using  (\ref{eq:postStreamLBTRT}) as $g^{\star}_i$.\label{algo:stream}
	\item Fill-in the remaining unknown populations $g^{\star}_i$ using interpolated bounce-back from (\ref{eq:IBB_LI}).
	\item Compute temporary populations moments $(\rho^{\star}, \bm{u}^{\star})$ from (\ref{eq:macroVarLB}).
	\item Then compute $g^{eq,{\star}} (\rho^{\star}, \bm{u}^{\star})$ at the fresh nodes from (\ref{eq:HeLuoEquilibroum}).
	\item Carry out local collision step for all fresh nodes using (\ref{eq:postCollLBTRT}).\label{algo:coll}
\end{enumerate}
\noindent
Steps (\ref{algo:stream})--(\ref{algo:coll}) are repeated for 5 inner-iterations, which has been shown to give sufficiently converged values of $(\rho^{\star}, \bm{u}^{\star})$ \citep{marson_EnhancedSinglenodeLattice_2021}.
During the inner-iterations, we update $\rho, \bm{u}$ and populations $g^{eq}, g$ only at the fresh nodes (and not at their link-wise neighbours).

\subsection{Numerical convergence}
\label{sec:methods:sub:convergenceSSF}
The computations are initiated with the fluid at rest in the channel. As the simulation advances in time, the continuous evolution of the instantaneous flow field is obtained.
At the end of every oscillation period $T$, we compute the developing SSF, $\bm{u}^{{ss},\bullet}$, by numerically integrating the LBM solution over all time steps of the last competed time period.
At the end of a time period, say when $t=T^{\prime}$, this can be written as,
\begin{equation}
	\label{eq:SSFcalc}
	\bm{u}^{{ss}, \bullet} (\bm{x})|_{T^{\prime}} = \frac{\Dt}{T} \sum_{i}^{T/\Dt}  \bm{u} (\bm{x}, T^{\prime}-T+i \Dt).
\end{equation}
\noindent
This developing SSF velocity field is then compared with the developing SSF field computed at the end of the previous time period, say when $T^{\prime \prime} = T^{\prime}-T$, using the $L2$ norm as,
\begin{equation}
	\label{eq:SSFL2norm}
	{E}_{L2} = \sqrt{  \frac{ \sum_{\bm{x}} \left( \bm{u}^{{ss},\bullet}(\bm{x})|_{T^{\prime \prime}} - \bm{u}^{{ss},\bullet}(\bm{x})|_{T^{\prime}} \right)^2 }{ \sum_{\bm{x}} \left( \bm{u}^{{ss},\bullet}(\bm{x})|_{T^{\prime \prime}} \right)^2 }}. 
\end{equation}
\noindent
When ${E}_{L2}$ falls below 1\%, the SSF is considered to have converged, and the simulation is set to terminate after running for one additional time period, $t = T^{\prime} + T$.  
During this final period, the time-periodic instantaneous velocity field $\bm{u}(\bm{x}, t)$ and the steady streaming flow $\bm{u}^{{ss}}(\bm{x})$ are recorded.

\section{Non-propagating contractions and steady streaming flow}
\label{sec:nonPropagating}

\begin{figure}
	\centering
	\includegraphics[width=\textwidth]{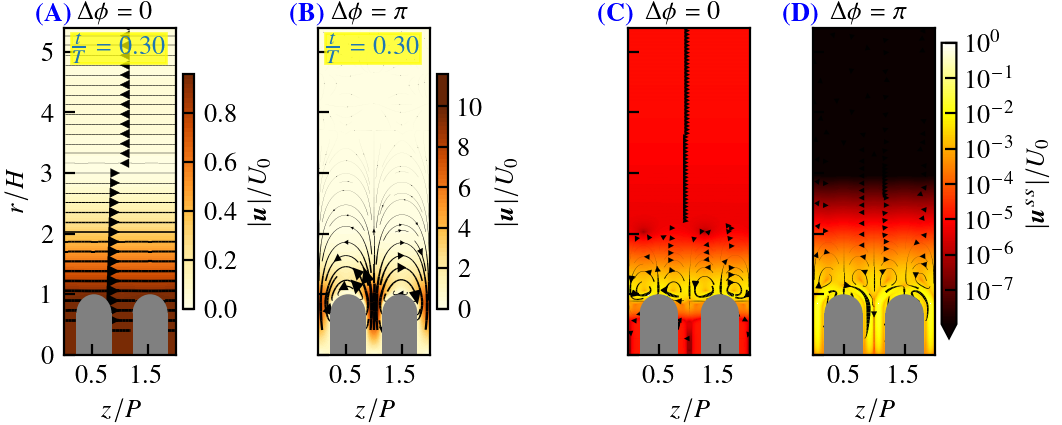}
	\caption{Flow induced by non-propagation oscillations of the villi over two neighbouring villi at $\Wo=0.5$. Instantaneous flow at $t/T=0.3$ for \textbf{(A)} in-phase ($\DPhi=0$) and \textbf{(B)} out-of-phase ($\DPhi=\pi$) villi motion.
	The colour fields map the magnitude of instantaneous velocity ($|\mathbf{u}|/U_0$).
	Time-averaged steady streaming flow-fields ($\bm{u}^{ss}$) for \textbf{(C)} in-phase ($\DPhi=0$) and \textbf{(D)} out-of-phase ($\DPhi=\pi$) villi motion.
	The colour fields map the magnitude of $|\mathbf{u}^{ss}|/U_0$.
	The width of the streamlines is proportional to the local velocity magnitude. See also figures S2-S4 in supplementary material.}
	\label{fig:instantAndSSF_Dphi_0_pi}
\end{figure}

In this section, we briefly tackle the limiting cases $\DPhi=0$ and $\pi$, which result in non-propagating contractions.
The flow velocity contours and streamlines presented in the remainder of the paper are shown for $\tilde{a}=0.2$, unless otherwise stated.
Flow velocities are rescaled by the characteristic villus velocity $U_0 = \omega a$, and plotted streamline width is proportional to the local velocity magnitude to illustrate flow strength.

\subsection{Instantaneous flow}
\label{sec:nonPropagating:sub:instant}

Figure~\ref{fig:instantAndSSF_Dphi_0_pi}(A) shows the instantaneous flow-field for $\DPhi = 0$, for $\Wo=0.5$ at a given time instance.
This is the 2D equivalent of the synchronised oscillations of an infinite array of villi recently studied in details by \cite{puthumana2022steady}.
At low $\Wo$, the flow closely follows the oscillations of the villus walls.  
As $\Wo$ increases and $\StLayer$ decreases, inertial effects progressively localize the flow near the villus tips and introduce a lag in the lumen (see figure S2).  
This results in transient recirculation zones above the villus tips.

When $\DPhi = \pi$, neighbouring villi oscillate exactly out of phase, as shown in figure~\ref{fig:instantAndSSF_Dphi_0_pi}(B) for $\Wo=0.5$ at a given time step. 
This motion generates counter-rotating vortical flow in the domain.
With increased $\Wo$, these vortical structures localize near villi tips (similarly to $\DPhi=0$ cases) due to radial confinement by a decreasing $\StLayer$ (see figure S3).

\subsection{Steady streaming flow}
\label{sec:nonPropagating:sub:SSF}
We compute the steady streaming flow-field (SSF) $\bm{u}^{ss}$ as the time-integrated mean flow, using (\ref{eq:SSFcalc}).
These are plotted for $\Wo=0.5$ in figures~\ref{fig:instantAndSSF_Dphi_0_pi}(C) and (D) for $\DPhi=0$ and $\DPhi=\pi$, respectively.
The SSF is characterized by counter-rotating vortices just above the villus tips, with amplitudes several orders of magnitude lower than the instantaneous flow velocity.
 
For the synchronised oscillation case $\DPhi=0$, an outer region of weak unidirectional steady flow develops in the lumen (see figure S4).
The strength of this flow increases with increasing $\Wo$ in agreement with the Rayleigh streaming theory \citep{puthumana2022steady, tatsuno1973circulatory}.
No such outer streaming region develops for the case $\DPhi=\pi$.
The key point here is that for non-propagating contractions, $\DPhi=0$ and $\pi$, irreversible SSF  manifests as an inertial phenomenon, and $\bm{u}^{ss}(\bm{x}) \approx 0$ in the Stokes flow regime when $\Wo \ll 1$

\begin{figure}
    \centering
    \includegraphics[width=\linewidth]{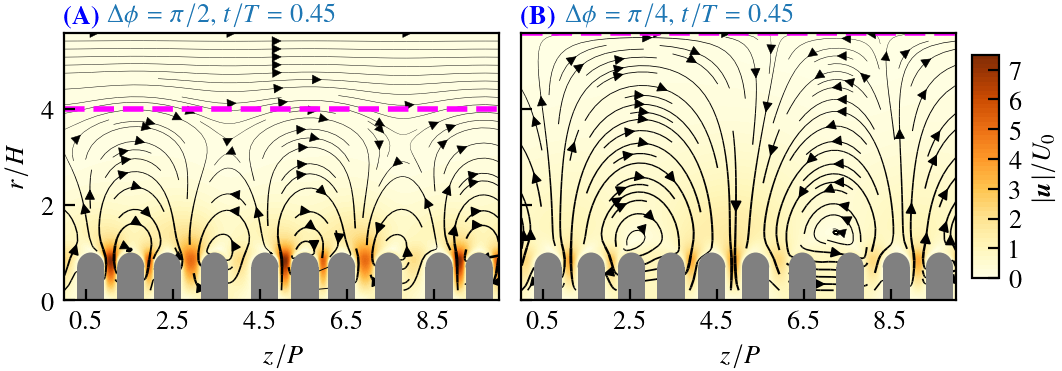}

    \caption{Snapshots of the instantaneous flow field for $\tilde{a}=0.2$ and $\Wo=0.16$, at $t/T=0.45$, for \textbf{(A)} $\DPhi = \pi/2$ and \textbf{(B)} $\DPhi = \pi/4$. The dashed (magenta) line marks the approximate separation between the mixing layer and the advected layer. Note the absence of the advected layer in \textbf{(B)}. See supplementary movies 1 and 2.}
	\label{fig:instantVel_lowWo}
	\vspace{0.5cm}
    \includegraphics[width=\linewidth]{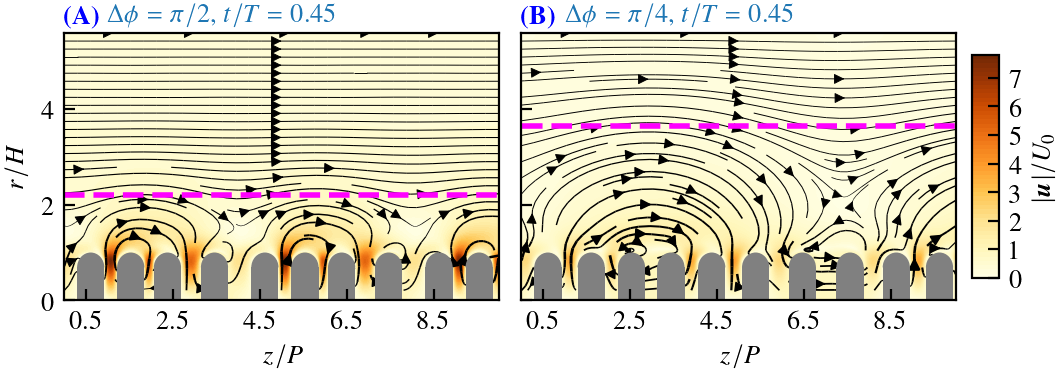}
    \caption{Snapshots of the instantaneous flow-field for  $\tilde{a}=0.2$ and $\Wo=2.82$, at $t/T=0.45$, for \textbf{(A)} $\DPhi = \pi/2$ and \textbf{(B)} $\DPhi = \pi/4$. The dashed (magenta) line marks the approximate separation between the mixing layer and the advected layer. See supplementary movies 3 and 4.} \label{fig:instantVel_highWo}

	\vspace{0.5cm}

    \includegraphics[width=\textwidth]{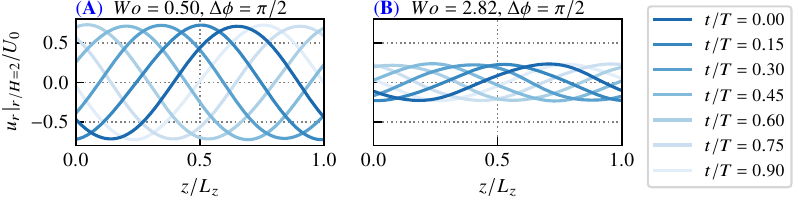}
	\caption{Evolution of the radial velocity component $u_r/U_0$, for $\DPhi=\pi/2$ and $\tilde{a}=0.2$, measured at height $r/H=2$. It is plotted for increasing time-fractions $0 \le t/T < 1$ (opaque to transparent) over the axial domain $z/L_z$, for \textbf{(A)} $\Wo=0.50$ and \textbf{(B)} $\Wo=2.82$. We see travelling velocity wave kinematics, moving from right to left ($-z$ direction) in both panels.}
	\label{fig:travelWaveVel}
\end{figure}

\section{Pendular-wave contractions and opposing steady streaming}
\label{sec:propagating}

In this section, we investigate the flow dynamics generated by the propagating contractions and relaxations of the intervillous spaces along $-\hat{\bm{z}}$, as illustrated in Figure~\ref{fig:schematic}(D).  
These correspond to the cases simulated with a phase lags in the range $0 < \DPhi < \pi$, for different $\Wo$.

\subsection{Instantaneous flow}
\label{sec:propagating:sub:instant}
Figures~\ref{fig:instantVel_lowWo} and \ref{fig:instantVel_highWo} show the  flow-fields for $\Wo=0.16$  and $\Wo=2.82$, respectively, for two values of $\DPhi$ in panels (A) and (B), at a given time instance.
From the figures, we notice the emergence of a boundary layer above the villi tips (dashed magenta line), that separates the flow into two distinct regions.
The velocity field between the villi (grey) and this boundary is characterised by asymmetric counter-rotating vortical flows of unequal strength.
These vortical flows originate in the contracting intervillous gaps, and terminate in those that are expanding.
We call this region the \emph{mixing boundary layer} (MBL) and its associated height is noted as $\ell$.
Above the MBL, until the central axis of the lumen at $r=R$, we observe near-uniform unidirectional flow moving along $+\hat{\bm{z}}$.
We term this region as the \emph{advected layer}.

Curiously, the vortical structures in the mixing layer are seen to propagate along $-\hat{\bm{z}}$, carried by the imposed propagating pendular-wave along the wall, while the axial flow in the advected layer goes opposite, along $+\hat{\bm{z}}$ (see supplementary movies 1--4).
For the case $\DPhi=\pi/4$ and $\Wo=0.16$ (figure~\ref{fig:instantVel_lowWo}(B)), the mixing layer extends all the way up to $r=R$, and the advected layer disappears.
In general, the height of the MBL increases with decreasing $\DPhi$, until it reaches the channel center-line.
Additionally, for a given value of $\DPhi$, the MBL height shrinks with an increase in $\Wo$, as can be observed when comparing figures~\ref{fig:instantVel_highWo} with \ref{fig:instantVel_lowWo}.
The decrease of $\ell$ with $\Wo$ is shown in supplementary movie 5.
For example, in the aforementioned case (at $\DPhi=\pi/4$), the MBL shifts from the center-line for $\Wo=0.16$ to approximately halfway between the center-line and villi tips for $\Wo=2.82$.

The effect of propagating wave along the villi-wall is felt throughout the lumen zone, irrespective of the MBL height.
Figure~\ref{fig:travelWaveVel} shows the measured radial velocity $u_r$ at $r=2H$ in the lumen, in panel (A) $\Wo=0.5$ and (B) $\Wo=2.82$, at increasing time-fractions $t/T$.
Right-to-left travelling wave kinematics is clearly seen in these plots, with a decrease in wave amplitude with increased $\Wo$.
A similar behaviour, but with a non-zero positive mean, is also seen for the axial velocity component $u_z$ (data not shown).
This behaviour of the instantaneous velocity is seen both in the mixing and the advected layer, and the wave amplitude is seen to monotonically decreases with increasing radial distance $r$. 

Whereas the propagating pendular-wave induces a travelling wave in the lumen (figure~\ref{fig:travelWaveVel}), the emergence of an advected layer that axially pumps fluid in the opposite direction is an intriguing phenomenon (figures~\ref{fig:instantVel_lowWo} and \ref{fig:instantVel_highWo}).  
This behavior is counter-intuitive in the context of momentum transport \textit{via} travelling waves, and contrasts with canonical peristaltic flow theory \citep{jaffrin1971peristaltic}, in which the wave transports the fluid in its direction of propagation.

\begin{figure}
	\centering
	\includegraphics[width = 1\textwidth]{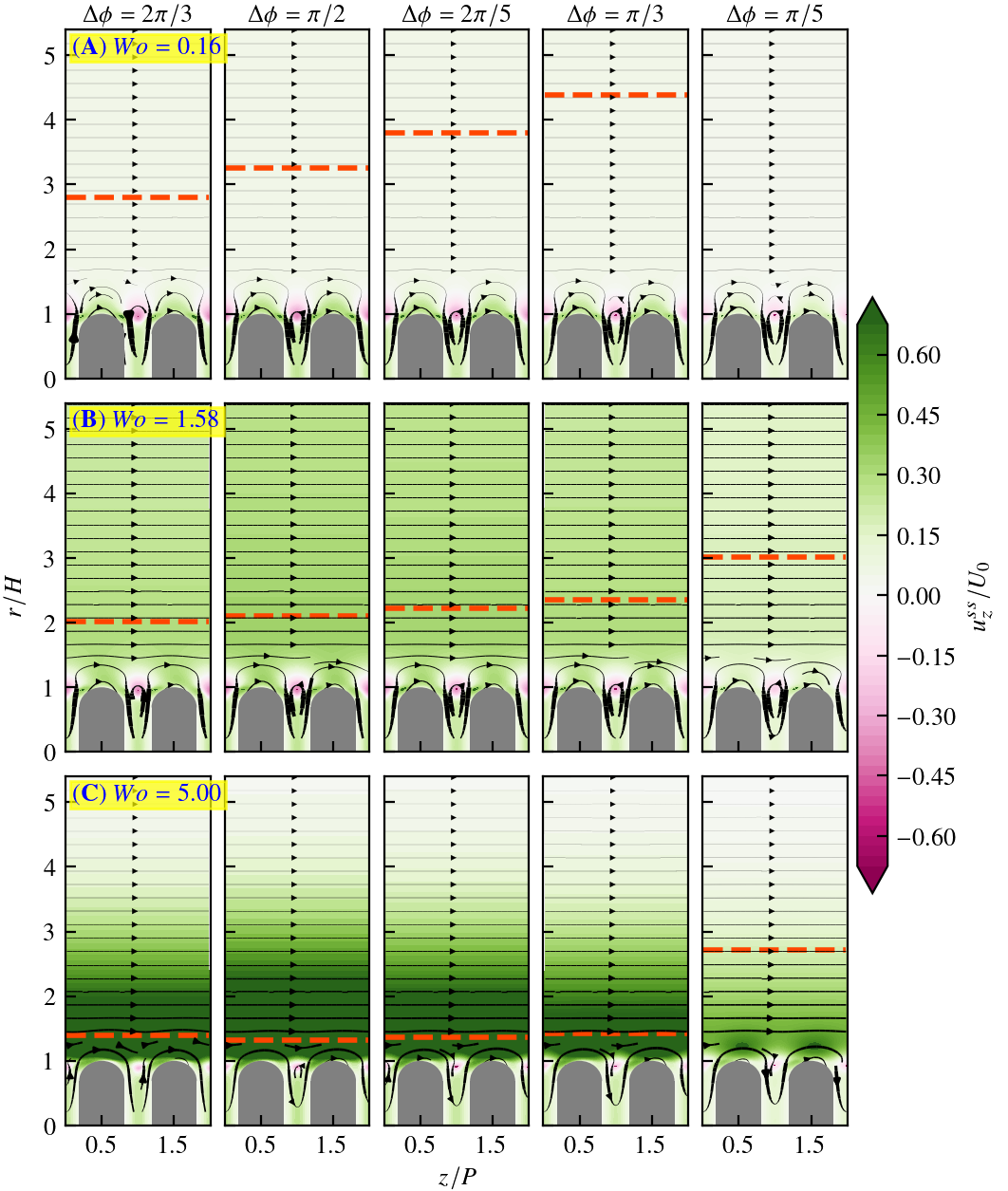}
	
	\caption{Steady streaming flow-field ($\bm{u}^{ss}$) streamlines for $\tilde{a}=0.2$,  plotted around a pair of adjacent villi for three increasing  Womersley numbers, \textbf{(A)} $\Wo=0.16$, \textbf{(B)} $\Wo=1.58$, and \textbf{(C)} $\Wo=5.0$ (row-wise). For each $\Wo$, panels for five decreasing $\DPhi$ values (column-wise) are shown. The colour-map plots the axial component of the steady streaming flow (SSF), ${u}^{ss}_z$. Note  that the SSF pattern shows axial periodicity over the intervillous distance $P$, for all cases. The dashed (orange) lines in each panel show the approximate mixing layer height $\ell$ seen in the respective instantaneous flow-fields.}
	\label{fig:SSF_map_propagating}
\end{figure}

\subsection{Steady streaming flow}
\label{sec:propagating:sub:SSF}
The axial fluid pumping effect is not limited to the advected layer, but occurs throughout the lumen including in the mixing layer.
This is demonstrated by the figure~\ref{fig:SSF_map_propagating} showing the time-integrated SSF computed for increasing $\Wo=0.16, 1.58$ and $5.0$, respectively.
For each $\Wo$ we show five panels for decreasing values of $\DPhi$ (column-wise).
The flow is shown over two neighbour villi (at their mean axial positions) in each simulated system, where the color-field plots the axial component $(u^{ss}_z)$ of the SSF velocity. 
Note that the SSF has a periodicity of $P$ in the axial direction. 
The dashed line in the plots shows the approximate height of the MBL seen in the instantaneous flow-field for each case.

For $\Wo=0.16$ and $1.58$, flow in the lumen is highly uniform in the $+r$ direction, except in and around villi zone.
Nearer the villi, flow becomes serpentine (along $+\hat{\bm{z}}$), descending down through the intervillous spaces and curving above the villi tips.
We see a small SSF vortex just above the intervillous space, as this serpentine flow meets the uniform flow above.
With an increase in $\Wo=5.0$, the flow is no longer radially uniform in the lumen and its strength gets greatly confined nearer the villi tips, with near zero net flow near the channel center-line.

It is worth noting that the MBL height $\ell$ does not appear to leave any signature on the SSF flow field, as evidenced by the dashed lines in figure~\ref{fig:SSF_map_propagating}.
We therefore cannot directly correlate the radial variations of the SSF patterns with the emergence of the MBL discussed in section~\ref{sec:propagating:sub:instant}.

\begin{figure}
	\centering
	\includegraphics[width=\textwidth]{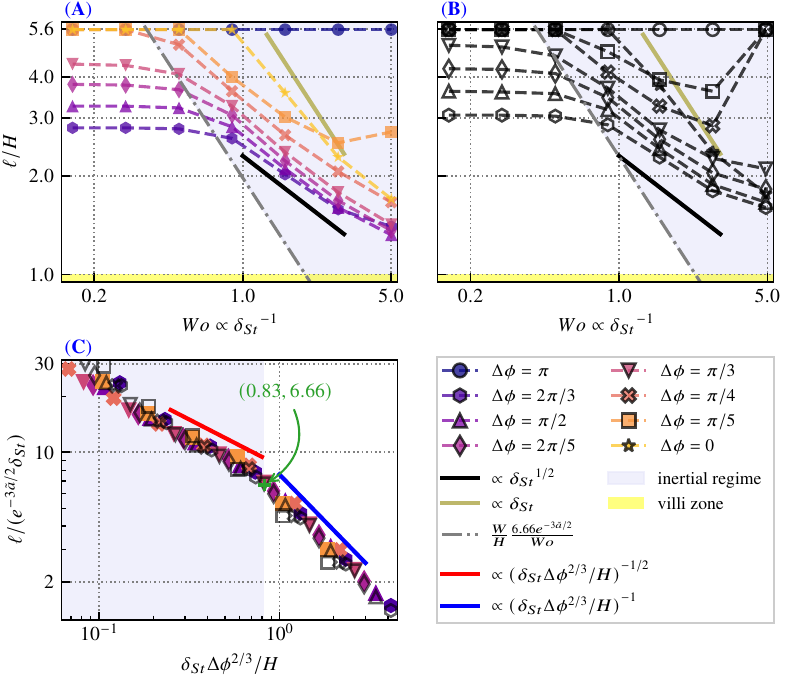}
	\caption{Plot of the mixing layer height $\ell$ against the Womersley number $\Wo = W / \StLayer$, for various $\DPhi$, for \textbf{(A)} $\tilde{a}=0.2$ and \textbf{(B)} $\tilde{a}=0.1$. 
	Note the plateauing of the curves when $Wo \le 1$, and their rapid decrease with increasing $\Wo$, when $\Wo \ge 1$, indicating a flow regime transition.
	The measured mixing layer heights $\ell$ from \textbf{(A)} and \textbf{(B)}, collapses as a double-power law when appropriately rescaled, as plotted in \textbf{(C)}, delineating the two regimes.
	Note that in \textbf{(C)} we have not show the data-points for $\DPhi = 0$ and $\pi$.
	In \textbf{(C)}, large $x$-axis shows the viscosity dominant regime, while at small $x$-axis shows the inertial regime.
	Contrast the scaling for $\ell$ in the inertial regime $(\propto \StLayer^{1/2})$ with that for the theoretical boundary layer ($\propto \StLayer$) for an oscillating flat plate \citep{schlichting_BoundaryLayerTheory_1960}.
	Transition between the two regimes appears to be smooth, and we visually identify (0.83,6.66) as the critical transition point in \textbf{(C)}.
	From this critical point, we obtain the functional transition curve: $\ell^c = 6.66 e^{-3\tilde{a}/2} W / \Wo$, which is plotted as dash-dotted (grey) lines in \textbf{(A)} and \textbf{(B)}.
	These lines indicate the viscous to inertial regime transition, for all $\DPhi$ and $\tilde{a}$.}
	\label{fig:mixingLengthEvol}
\end{figure}

\section{Mixing boundary layer evolution}
\label{sec:MBL}

In the previous section, simulations revealed a separation of the flow field into a mixing boundary layer above the villi and an advective layer in the center of the lumen. The emergence of an viscous boundary layer, normal to planar boundaries, is a classical phenomenon in pulsating flows at finite inertia \citep{schlichting_BoundaryLayerTheory_1960, loudon_UseDimensionlessWomersley_1998}. 
This is typically referred to as the Stokes boundary layer, and scales as $\StLayer \sim \sqrt{\mu / (\omega \rho)}$.
In this section, we show that patterning the channel with villi leads to departure from the classical theory.

\subsection{Mixing boundary layer height}
\label{sec:MBL:sub:MBLheight}

The MBL height $\ell$ separates the vortical flow region from a unidirectional flow region (advected layer) above it.
In our simulations, the axial velocity in the advected layer is always positive, i.e.\ $u_z > 0$.
We use this fact to characterize the height of MBL, by first decomposing the instantaneous axial velocity according to its direction as,
\begin{equation}
	\label{eq:plusMinusVel}	
	{u}^-_{z} (r,z,t) = \begin{cases}
		{u}_{z}, & \text{if } ~ {u}_{z} < 0\\
		0, & \text{otherwise},
	\end{cases}\quad \text{and}\quad
	{u}^+_{z} (r,z,t) = {u}_{z} - {u}^-_{z}
\end{equation}
and then computing the time and axial average of the negative velocity $u^-_{z}$ as,
\begin{equation}
	\label{eq:minusVelAveraging}
	\langle u^{ss-}_z \rangle_z (r) = \frac{1}{L_z T} \int_{0}^{L_z} \int_{t}^{t+T} u^-_z (t,z,r) dt dz,
\end{equation}
which depends only on the radial distance $r$.
We now identify the MBL height $\ell$ as the shortest radial distance measured from the villi base at which the $\langle u^{ss-}_z \rangle_z (r \coloneq \ell) = 0$. With this definition we obtain the MBL height $\ell$, measured from the villi base,  within which the instantaneous velocity field is characterized by significant negative axial flow $u^{-}_z (z, r, t)$, and above which the $u^{-}_z (z, r, t)$ is negligible.

We measure $\ell$ for all simulated cases, and its non-dimensional value is plotted against increasing $\Wo$ in figure~\ref{fig:mixingLengthEvol} for various $\DPhi$, in panels (A) for $\tilde{a}=0.2$ and (B) for $\tilde{a}=0.1$.
We first look at the non-propagating case with $\DPhi=0$ (star marker) and $\DPhi = \pi$ (circle marker).
For synchronised oscillations of the villi, $\DPhi=0$, $\ell$ remains constant at the maximum radial distance of $r=R$ for small $\Wo$, and then sharply decreases with a slope $\sim \StLayer$ at increased $\Wo$.
This linear decrease of $\ell$ with increased $\Wo$, mimics the dynamics due to an oscillating flat plate, except at $\Wo=5.0$, where structure of the villi-wall boundary causes deviations.
Here, the mixing boundary layer is controlled by the Stokes layer, and the effect of the villi-wall can be approximated by that of an oscillating flat plate positioned at an offset to the villi-wall base.
\cite{puthumana2022steady} have shown that the mixing layer for synchronous oscillations is entirely an inertial phenomenon and can be controlled by changing the intervillous confinement.
On the other hand, when $\DPhi=\pi$, we have neighbour villi that oscillate exactly out of phase and generate a mixing layer.
This mixing layer extends up to the channel centerline ($r = R$), without the emergence of an advected layer.

A distinct class of behaviour is seen in the MBL $(\ell)$ plots for propagating contractions, $0 < \DPhi < \pi$.
Here, $\ell$ plateaus to a constant value at low $\Wo$, which depends solely on $\DPhi$ and $\tilde{a}$.  
This is a departure from the classical oscillating boundary layer theory \citep{ schlichting_BoundaryLayerTheory_1960}, which predicts that at low $\Wo$ (i.e.\ in the Stokes flow regime), the viscous dissipation layer should encompass the entire channel.
The plateau value of $\ell$ decreases as $\DPhi$ increases from $\pi/5$ to $2\pi/3$.

At high $\Wo$ in the inertial regime, $\ell$ decreases with a slope $\sim \sqrt{\StLayer}$, contrasting with the linear scaling seen for $\DPhi=0$. 
Comparing figures~\ref{fig:mixingLengthEvol}(A) and (B), we also see that decreasing $\tilde{a}$ causes an increase in the relative levels of the measured MBL height.
At smaller $\DPhi$, e.g.\ for $\DPhi=\pi/5$ (square markers), the Stokes to inertial transition of $\ell$ occurs at a smaller $\Wo$, as compared to that at larger values e.g.\ $\DPhi = \pi/2$ (triangle markers).
The critical transition $\Wo$ is smaller for larger plateau values of $\ell$ (i.e.\ smaller $\DPhi$ values).
Here, for $\DPhi \le \pi/4$, the increasing mixing layer height is truncated by the radial confinement of the geometry.
For these cases, the mixing layer extends across the entirety of the lumen at $\Wo \ll 1$, and the advected layer disappears.

The plots also appear to show an apparent increase of $\ell$ at high $\Wo=5.0$, for certain cases, namely for $(\tilde{a}=0.2, \DPhi=\pi/5$) (see supplementary movie 6), and for  $(\tilde{a}=0.1, \DPhi=\pi/5)$ and $(\tilde{a}=0.1, \DPhi=\pi/4)$ (see supplementary movie 7).
Upon further investigation, we find that the flow structure in these cases is unlike that observed in the rest of the simulations.
Here, vortical flow pattern emanating from the villi is altered, with the smaller counter-clockwise vortical flow assuming a distinct hourglass-like shape, while the larger clockwise vortical flow reaching the channel center.
Despite the high oscillatory inertia at $\Wo = 5.0$, we have a mixing layer height that reaches the channel center.
Here, the steady streaming flow gets significantly weakened (see figure~\ref{fig:SSF_map_propagating}, panel: $\DPhi=\pi/5, \Wo=5.0$) and the vortical instantaneous flow dominates throughout the luminal space.
This changed flow structure indicates the existence of another inertial flow transition beyond $\Wo \gtrsim 5$, which would also depends on $\DPhi$ and $\tilde{a}$.
Here the computed averaged MBL height ($\ell$) is inconsistent due to the change in the flow structure for these cases.
In this paper we limit ourselves to $\Wo \le 5.0$, and a well defined $\ell$, and therefore exclude these three points from further analysis in the present section.

The mixing layer height data shows a phenomenological collapse onto a double power-law master curve, as shown in figure~\ref{fig:mixingLengthEvol}(C).
Since the data for non-propagating cases ($\DPhi=0$ and $\pi$) exhibits a different class of behaviour, it does not collapse onto the master curve, and is therefore not shown.
The $x$ variable for the collapse is $\StLayer \DPhi^{2/3} / H$ and the $y$ variable is $\ell / (e^{-3\tilde{a}/2} \StLayer)$.
Physically, the $y$ variable represents the MBL height normalized by the viscous penetration layer above the villi, after correcting for spatial attenuation associated with the oscillatory amplitude.
The $x$ variable is $\propto 1/\Wo$ and quantifies the relative reduction of oscillatory inertial effects, rescaled by the leading-order dependence of the MBL height on the phase lag.
For large values of the $x$ variable, the flow is dominated by viscous effects, leading to the power law relation $y \propto x^{-1}$ (blue line), and thus $\ell \propto \DPhi^{-2/3} e^{-3\tilde{a}/2}$.  
Here, the mixing layer height $\ell$ is unaffected by oscillatory fluid inertia and is therefore independent of $\StLayer$, identifying the parameter set that falls within the viscous dominated regime.

\begin{figure}
	\centering
	\includegraphics[width=\textwidth]{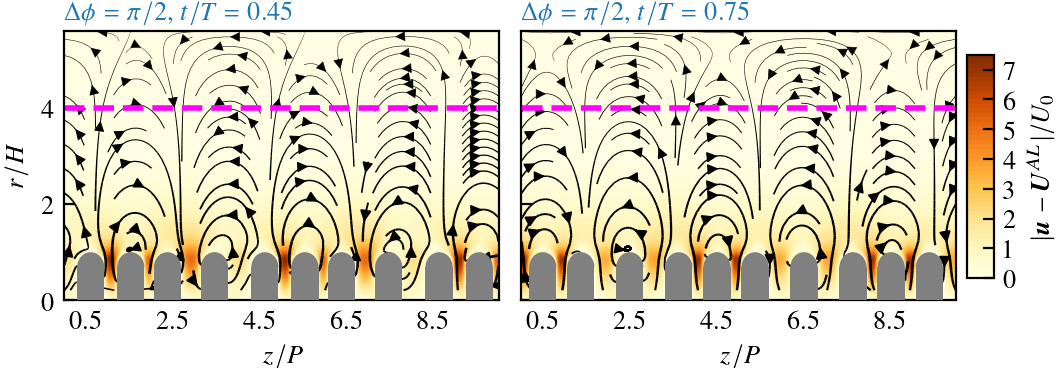}
	\caption{Snapshots of the purely oscillatory velocity field, visualized by subtracting a constant velocity $(U^{AL}_z, 0)$ from the instantaneous velocity $\bm{u}$, shown for two time-fractions $t/T=0.45$ and $0.75$, for the case $\DPhi = \pi/2$, $\Wo=0.16$ and $\tilde{a}=0.2$. $U^{AL}_z$, computed from (\ref{eq:advectingLayerVelScale}), is a measure of the irreversible axial flow velocity in the advected layer. The dashed (magenta) separation line between the mixing and advected layers is reproduced from figure~\ref{fig:instantVel_lowWo}\textbf{(A)}.}
	\label{fig:subtractedVorticalFlow}
\end{figure}

\begin{figure}
	\centering
	\includegraphics[width=\textwidth]{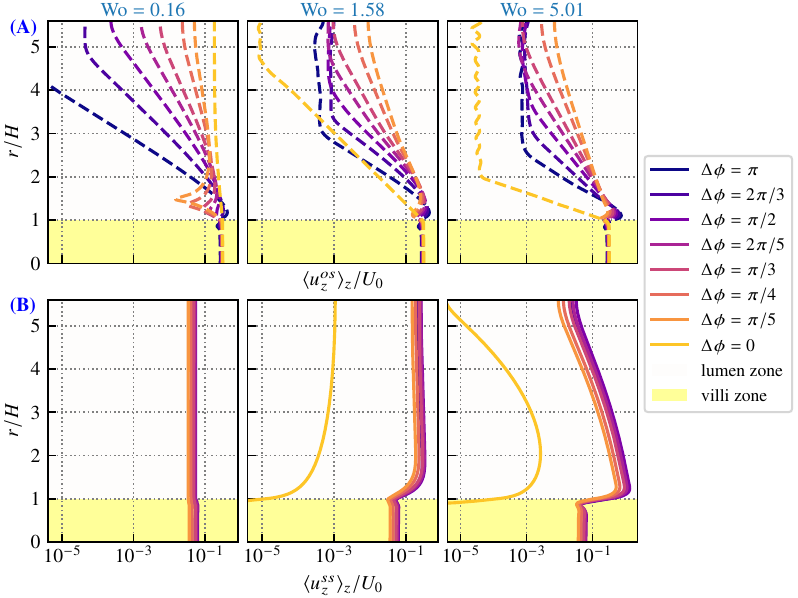}
	\caption{Axially averaged  $(\langle \cdot \rangle_z)$ $z$-components of \textbf{(A)} the oscillatory axial velocity $\langle u^{os}_z \rangle_z$  and \textbf{(B)} the steady streaming axial velocity $\langle u^{ss}_z \rangle_z$.
		These measures are computed as functions of radial distance $r/H$, from equations~(\ref{eq:SSFvel_axAvg}) and (\ref{eq:OSCvel_axAvg}) for various $\DPhi$ and plotted (column-wise) for three increasing Womersley numbers $\Wo = 0.16, 1.58$ and $5.01$. $\tilde{a}=0.2$.
		The plots demonstrate that the oscillatory flow decays exponentially in $r$, while the steady flow (which is mostly uniform for low $\Wo$) does not.}
	\label{fig:axialAvgVel_oscAndSteady}
\end{figure}

On the other hand, as the $x$ variable decreases, the flow transitions into the oscillatory inertial regime, with the power law relation $y \propto x^{-1/2}$ (red line).
The mixing layer height now depends on $\Wo$ and scales as $\ell \propto \StLayer^{1/2} \DPhi^{-1/3}  e^{-3\tilde{a}/2}$.
The double power law collapse therefore shows a transition from the Stokes regime (large $x$) into the inertial regime (small $x$), with a relatively smooth transition in between.
At smallest values of $x$ variable, we see that the data collapse starts to fray, suggesting that this flow structure will not sustain at highly elevated oscillatory inertia.
We visually identify the critical transition point between the two regimes as $(x\approx0.83$, $y\approx6.66)$. 
This allows us to deduce a critical transition line expressed in terms of $\DPhi$ as  
$\ell^c = 5.5 e^{-3\tilde{a}/2} H \DPhi^{-2/3}$,  
and similarly, in terms of $\Wo$, as  
$\ell^c = 6.66 e^{-3\tilde{a}/2} W / \Wo$.
This Stokes-to-inertial critical transition line is overlaid on the unscaled $\ell/H$ data in figures~\ref{fig:mixingLengthEvol}(A) and (B) (grey dash-dotted line).
The inertial regime is shown by the grey shaded region in all plots in the figure~\ref{fig:mixingLengthEvol}.

\subsection{Competing oscillatory and steady flows}
We now look into the physical mechanism that leads to the appearance of distinct  mixing and advected layers.
Towards this end, we subtract the characteristic flow velocity $\bm{U}^{AL}$ of the advected layer from the entire flow-field.
Here, the axial component of the characteristic velocity is computed as,
\begin{equation}
	\label{eq:advectingLayerVelScale}
	U^{AL}_z = \frac{1}{H L_z} \int_{0}^{L_z} \int_{5H}^{R} u_z dr dz,
\end{equation}
making sure that we integrate well above the MBL height, and the radial component is set to $U^{AL}_r = 0$.

Figure~\ref{fig:subtractedVorticalFlow} shows this shifted velocity field for $\DPhi=\pi/2$ and $\Wo=0.16$, with the height of the MBL reproduced (dashed magenta line) from figure~\ref{fig:instantVel_lowWo}(A).
The shifted velocity field consists entirely of counter-rotating, semi-vortical flow structures that now extend up to the central symmetry boundary.
This structure is similar to the one seen when the mixing layer extends till the center line (e.g.\ figure~\ref{fig:instantVel_lowWo}(B)).
This pattern is produced irrespective of $\DPhi$ in the Stokes flow regime ($\Wo=0.16$), demonstrating that the total flow-field is simply a sum of the axial steady flow and oscillatory components.

The distinct mixing and advected layers seen in the simulations manifest due to the competition between the strength of the oscillatory flow, which generates the counter-rotating vortical structures, and the axial steady flow shown in figure~\ref{fig:SSF_map_propagating}.
In order to understand the MBL height $\ell$ behaviour, we map the radial dependence of the steady (irreversible) and vortical flows as,
\begin{align}
	\label{eq:SSFvel_axAvg}
	\langle u^{ss}_z \rangle_{z} (r) &= \frac{1}{L_z} \int_0^{L_z} u^{ss}_z (z,r) dz \\
	\label{eq:OSCvel_axAvg}
	\langle u^{os}_z \rangle_{z} (r) &= \frac{1}{2T L_z} \int_0^{L_z} \int_t^{t+T} |u_z (t, z,r) - u^{ss}_z (z,r) | dt dz,
\end{align}
where, $\langle u^{ss}_z \rangle_{z}$ and $\langle u^{os}_z \rangle_{z}$ are the axially averaged steady streaming and oscillatory axial velocities.
Figure~\ref{fig:axialAvgVel_oscAndSteady} compares the strength of these two velocity measurements, for three increasing values of $\Wo$ (column-wise panels) and all $\DPhi$.
In figure~\ref{fig:axialAvgVel_oscAndSteady}(A), the oscillatory velocity $\langle u^{os}_z \rangle_{z}$ shows an exponential decay with increasing $r/H$, for various $\DPhi$.
This decay becomes more pronounced as $\Wo$ increases.
In contrast, as seen in figure~\ref{fig:axialAvgVel_oscAndSteady}(B), the steady streaming $\langle u^{ss}_z \rangle_{z}$ remains nearly constant, starting from the villi tips to the centre of the lumen at lower $\Wo$, while becoming increasingly bulged just above the villi tips at higher $\Wo$.

In summary, the pendular-wave generates a steady  axial flow.
This flow is nearly uniform, and depends only on $\DPhi$ when viscosity dominates over oscillatory inertia.
The propagating contraction-expansion kinematics of the villi-wall also generate axially oscillating flow, seen in the form of counter-rotating vortical structures.
The strength of this oscillatory flow decays exponentially along the radial direction, with a decay rate dependent on $\DPhi$.
The height of the MBL is so determined by a competition between these two effects.
Within the mixing layer the oscillatory axial flow predominates, while in the advected layer its strength falls below that of the steady streaming axial flow.
Hence, in viscosity dominated cases, $\ell$ depends only on $\DPhi$.
As oscillatory inertia increases, a decreasing Stokes layer $\StLayer$ enforces a dynamic radial confinement on both the steady and oscillatory flows, causing a depletion of the steady axial flow near the channel center, as well as a faster radial decay of the oscillatory axial flow.
This in turn causes the MBL height to decrease non-trivially in the inertial regime, as a function of both $\Wo$ and $\DPhi$, unlike that for a simple oscillating flat plate.

\section{Origin of irreversible axial flow opposing wave travel}
\label{sec:fluxesScaling}

As stated previously, the existence of irreversible flow in low and moderate $\Wo$  in a direction opposite to that of the wave travel contrasts with canonical peristaltic flow theory \citep{jaffrin1971peristaltic}.
In this section we aim to identify the physical mechanism responsible for the steady streaming (irreversible) axial flow opposing the pendular-wave propagation direction, as well as the instantaneous radial pumping.
The steady flow component is responsible for the emergent advected layer, as seen in the previous section.  
To this end, we develop scaling laws that predict the magnitude and direction of the irreversible axial fluxes and radial pumping, and test their validity against numerically integrated measures from the simulation results.

\subsection{Scaling laws}
\label{sec:fluxesScaling:subsec:scalingLaws}

At first glance, the irreversible axial flow pumping generated by the villi wall studied here is reminiscent of that observed in propagating waves along dense oscillating ciliary arrays \citep{ishikawa_FluidDynamicsSquirmers_2024, khaderi_MicrofluidicPropulsionMetachronal_2011, ding_MixingTransportCiliary_2014, hall_MechanicsCiliumBeating_2020}. 
Specifically, those undergoing \textit{antiplectic metachronal waves}, where the direction of wave propagation is opposite to that of net fluid pumping by the ciliary array \citep{ding_MixingTransportCiliary_2014}.
However, individual cilia possess are deformable (an additional degree of freedom) and trace non-reciprocal effective and recovery strokes during their oscillatory cycle. 
This breaks the time-symmetry of the Stokes equations, and generates irreversible fluid transport. 
This is clearly not the case for villi, which are rigid structures \citep{lim2014determination} and undergo reversible harmonic oscillations.
The non-reciprocal motion of the villi-wall, resulting in steady streaming axial flow even in the Stokes flow regime, is therefore more subtle.

Building on the \textit{Scallop Theorem} \citep{purcell1977life,lauga_LifeScallopTheorem_2011}, \citet{najafi_SimpleSwimmerLow_2004} proposed a minimal model for a swimmer in Stokes flow, consisting of three spheres linked in-line, where the inter-sphere gaps undergo a cyclic non-reciprocal sequence of contraction and extension. 
While the individual spheres follow reciprocal trajectories, the collective behaviour of the system breaks time-symmetry and generates net propulsion in Stokes flow regime \citep{najafi_SimpleSwimmerLow_2004, najafi_PropulsionLowReynolds_2005}.
Referring to figure~\ref{fig:schematic}(D), it becomes clear that a similar non-reciprocal sequence is executed by any two adjacent intervillous gaps (and their bounding villi) in our system. 
The villi wall can thus be viewed as an array of such non-reciprocally contracting and relaxing gaps, where the additional degree of freedom is introduced via the spatial phase lag $\DPhi$.
This phase-lagged actuation implies that each downstream (along $+z$) intervillous gap reaches its minimum or maximum size slightly ahead of its upstream neighbour. This sequence is never reversed in time, ensuring non-reciprocal boundary kinematics.

The contraction-expansion of the intervillous gaps pumps fluid radially, as illustrated by the arrows in figure~\ref{fig:schematic}(B). 
The non-reciprocal motion of adjacent intervillous gaps facilitates the net transfer of some parcels of this fluid from an upstream gap to its downstream neighbour.
The villi-wall boundary therefore irreversibly pumps fluid in the axial direction.
Based on this mechanistic description, we develop a scaling laws to describe the emergence of the oscillatory radial flow. 
The rate of expansion and contraction of the gap between two adjacent villi, denoted $\zeta_i$, along the $\hat{\bm{z}}$ direction is given by,

\begin{equation}
	\label{eq:gapExpansionRate}
	\dot{\zeta}_i = 2 a \omega \sin \left( \frac{\DPhi}{2} \right) \cos \left ( \omega t + \frac{\DPhi}{2} (2i-1) \right).
\end{equation}

Integrating this equation allows us to scale the peak oscillatory radial flux per intervillous gap, denoted $Q_r^{os,\max}$ (see Appendix~\ref{secX:geometricScalings:sub:radialFluxOsc} for details), which leads to,

\begin{equation}
\label{eq:scalingRadial}
	Q_r^{os,\max} = \frac{2 a \omega H}{\pi} \sin\left(\frac{\DPhi}{2}\right),
\end{equation}

Along the axial direction, the fluid trapped in the intervillous gaps undergoes non-reciprocal motion. Taking into account the coupled effect of the gap size and the fluid velocity within the gap, we get the direction and magnitude for the steady streaming axial flux averaged over the domain length $L_z$ (see Appendix~\ref{secX:geometricScalings:sub:AxialSSFluxVilli}). 
The flux within the villi zone is denoted $Q^{ss,v}_z$ and that within the lumen zone is denoted $Q^{ss,l}_z$, are scaled as,

\begin{equation}
\label{eq:scalingAxial}
	Q^{ss,v}_z = \frac{a^2 \omega H}{2 P} \sin \left( \DPhi \right) \quad \text{and} \quad Q^{ss,l}_z = \frac{a^2 \omega (R-H)}{2 P} \sin \left( \DPhi \right).
\end{equation}

\begin{figure}
	\centering
	\includegraphics[width=\textwidth]{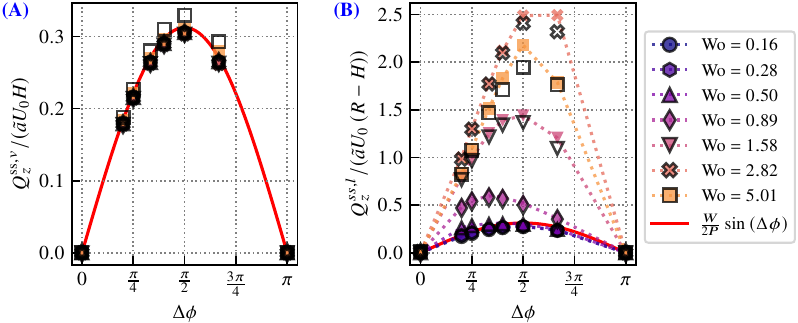}
	\caption{Axial steady-streaming fluxes integrated over \textbf{(A)} the villi-zone $Q_z^{ss,v}$ and \textbf{(B)} over the lumen-zone $Q_z^{ss,l}$, plotted against the phase lag $\DPhi$, for increasing $\Wo$.
	Flux quantities are defined as averages taken over the channel length $L_z$.
	Filled markers (coloured) plot simulations with $\tilde{a}=0.2$ and empty markers (black) with $\tilde{a}=0.1$.
	In both plots, the solid (red) curve is from equation \ref{eq:scalingAxial}, and  derived from geometric considerations detailed in section~\ref{secX:geometricScalings:sub:AxialSSFluxVilli}.
	Note that in \textbf{(B)} $Q_z^{ss,l}$, shows non-monotonicity with increasing $\Wo$, with the maximum irreversible steady flux occurring when $\Wo \approx 2.82$.}
	\label{fig:axialFluxesSSF}
\end{figure}

\subsection{Axial pumping}
\label{sec:fluxesScaling:subsec:axialFlux}

To confirm our mechanistic scaling arguments, we numerically measure the irreversible axial flux, $Q^{ss}_z$, separately in the villi zone ($0 \le r \le H$) and in the lumen zone ($H < r \le R$) averaged over the channel length $L_z$ as,

\begin{equation}
	\label{eq:fluxAxialSS_villiZone}
	Q^{ss,v}_z = \frac{1}{L_z} \int_{0}^{H} \int_0^{L_z} u^{ss}_z (z,r) dz dr 
\end{equation}
and 
\begin{equation}
	\label{eq:fluxAxialSS_lumenZone}
	Q^{ss,l}_z = \frac{1}{L_z} \int_{H}^{R} \int_0^{L_z} u^{ss}_z (z,r) dz dr.
\end{equation}

Figures~\ref{fig:axialFluxesSSF}(A) and (B) show the domain averaged axial steady streaming fluxes in the villi and lumen zones, respectively, for various $\Wo$.
For the irreversible SSF flux in figure~\ref{fig:axialFluxesSSF}(A), we see a remarkable data collapse on the scaling equation (\ref{eq:scalingAxial}) (solid red curve).
Here, the data begins to show deviations due to increasing inertial effects only for high  $\Wo=5.0$ for $\tilde{a}=0.1$.
The measured data for $Q^{ss, v}_z$ shows minimal dependence on $\Wo$, indicating that the flow in the villi zone remains well-approximated by the Stokes regime, even at higher $\Wo$.
This behaviour arises from the significant geometric confinement of the flow within the intervillous gaps, characterized by a small $(P-W)/H$.

The irreversible luminal flux is plotted in figure~\ref{fig:axialFluxesSSF}(B), rescaled appropriately by $\tilde{a} U_0 (R-H)$, as a function of $\DPhi$.
In the viscous-dominated flow regime at low $\Wo \ll 1$, the steady-state axial flux data, $Q^{ss, l}_z$, collapses onto the same functional form (solid red curve) given by (\ref{eq:scalingAxial}).  
The data-points here for $\Wo \le 0.28$ fall slightly below the scaling equation (solid line).
Thus, in the Stokes flow regime, we can approximate the total steady streaming axial flux across any height $h$ of the channel as $Q^{ss}_z = Q^{ss, v}_z + Q^{ss, l}_z \approx \left(W h \sin\left( \DPhi \right) \right ) / \left(2PH\right)$.

However with an increase in oscillatory inertia, the flux data no longer shows a functional collapse.
The flux here shows a non-monotonic behaviour with increasing $\Wo$ in figure~\ref{fig:axialFluxesSSF}(B).
Note that $Q^{ss, l}_z$ at $\Wo=5.0$ is lower that that at $\Wo=2.82$, for all $\DPhi$.
Alt increasing $\Wo$ increases the axial velocity of the fluid in the lumen, the increased inertia also limits the radial height to pump this fluid due to dynamic radial confinement of the flow near the villi tips (see figure~\ref{fig:SSF_map_propagating}).
The flux therefore reaches a maximum  at $\Wo \approx 2.8$ and falls beyond $\Wo > 2.82$.

\begin{figure}
	\includegraphics[width=\textwidth]{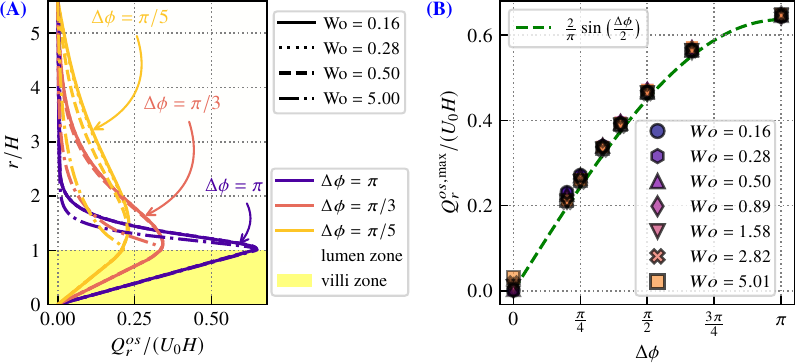}
	\caption{\textbf{(A)} The oscillatory radial flux $Q^{os}_r$ per intervillous gap evaluated from (\ref{eq:radialFluxPerN_OSC}) is plotted (on $x$-axis) against radial distance $r/H$ (on $y$-axis) for $\tilde{a}=0.2$ and three different $\DPhi=\pi$, $\pi/3$ and $\pi/5$.
	For each $\DPhi$, curves for four increasing Womersley numbers $\Wo=0.16$ (solid), $0.28$ (dotted), $0.5$ (dashed) and $5.0$ (dash-dotted) are shown.
	Note that radial flux curves  at low $\Wo$ (for $\Wo=0.16$ and $0.28$) coincide exactly. 
	Minor deviations from these coincident curves are seen with the onset of oscillatory fluid inertia, when $\Wo=0.5$, and large deviations are seen at increased oscillatory inertia for $\Wo=5.0$.
	\textbf{(B)} The peak value of radial oscillatory flux ($Q_r^{os,\max}$) per intervillous gap plotted against the phase lag $\DPhi$, for increasing $\Wo$. 
	Filled markers (coloured) are for $\tilde{a}=0.2$ and the empty markers (black) for  $\tilde{a}=0.1$. The peak flux data-points collapse solely as a function of $\DPhi$, when appropriately rescaled. 
	The simple scaling function is shown with dashed (green) line.}
	\label{fig:radialFluxOsc}
\end{figure}

\subsection{Radial pumping}
\label{sec:fluxesScaling:subsec:radialFlux}
The radial pumping of fluid in the channel cannot lead to an irreversible radial flow due to the confining symmetric top and bottom channel walls.
Instead, all of the radially pumped fluid recirculates, forming counter-rotating semi-vortical flow structures, as seen in figures~\ref{fig:instantVel_lowWo} and \ref{fig:instantVel_highWo}.

Radial pumping is quantified using radial oscillatory flux.
Due to radial flow confinement, mass conservation ensures that a simple integration of the velocity field along the $z$-axis yields zero net flux.
We therefore define the oscillatory radial flux per intervillous gap  as,
\begin{equation}
	\label{eq:radialFluxPerN_OSC}
	Q^{os}_r (r) = \frac{1}{2 T N} \int_{0}^{L_z} \int_{t}^{t+T} |u_r (t,z,r) - u^{ss}_r (z,r)| dt dz .
\end{equation}
Here, $N = 2 \pi / \DPhi $ is the number of intervillous gaps along axial periodic length $L_z = N P$.
Subtracting $u^{ss}_r (z,r)$ from the instantaneous velocity ensures that we capture the periodically oscillating component of the flux.
Unlike in the axial case, the radial flux per intervillous gap, $Q^{os}_r$, is defined to retain its radial dependence and to have equal magnitude in $\pm r$ directions.

Figure~\ref{fig:radialFluxOsc}(A) shows the non-dimensional oscillatory radial flux $Q^{os}_r (r)$ on the $x$-axis, as a function of the radial distance along $y$-axis, for three values of $\DPhi=\pi$, $\pi/3$ and $\pi/5$, at $\tilde{a}=0.2$.
The flux curves are plotted for four increasing values of $\Wo=0.16$ (solid), $0.28$ (dotted), $0.5$ (dashed) and $5.0$ (dash-dotted).
The radial flux rises to a peak value just outside the villi tips and then falls to zero at large $r$ for all cases.
Unsurprisingly, we observe that the maximum radial flux is obtained for $\DPhi = \pi$, corresponding to the case where the intervillous spaces experience the largest strain for a given villus displacement $a$. As $\DPhi$ decreases, the peak radial flux value also decreases. Moreover, the radial distance over which the flux decays to zero increases with decreasing $\DPhi$; the flux decays most rapidly for $\DPhi = \pi$ and most gradually for $\DPhi = \pi/5$.
This is consistent with the behaviour of the mixing layer height $\ell$ with decreasing $\DPhi$.

In the lumen zone, when $\Wo \ll 1$, (i.e.\  $\Wo=0.16$ and $0.28$), the flux curves are coincidental for all $\DPhi$.
This overlap demonstrates that in the absence of fluid inertia, the $Q_r^{os}$ and its radial dependence is solely controlled by the geometric motion of the villi-wall, set by the phase lag $(\DPhi)$.
As $\Wo$ increases ($\Wo \approx 0.5$), slight deviations away from the coincidental curves appear along $r$, indicating an onset of inertial effects.
In the oscillatory inertia dominated regime ($\Wo=5.0$), $Q_r^{os}$ curves decay much faster to zero along $+r$ in the lumen.
This again is a consequence of decreasing $\StLayer$, and the radial flux in the lumen is now a function of both $\DPhi$ and $\Wo$.
However, in the villi zone the radial flux profiles are governed solely by the phase lag $\DPhi$, and these always overlap irrespective of $\Wo$.
Thus, similar to the axial flux, the oscillatory radial flux in the villi zone remains unaffected by oscillatory fluid inertia due to greater geometric flow confinement.

The peak value of oscillatory radial flux along $r$ is identified as $Q_z^{os, \max} (r^{\max}) = \max(Q_z^{os}(r))$.
Figure~\ref{fig:radialFluxOsc}(B) plots the rescaled oscillatory radial flux as a function of $\DPhi$, for increasing $\Wo$.
Here, we see a remarkable data collapse onto the scaling form given by equation (\ref{eq:scalingRadial}) (dashed green line), for all $\Wo$ and $\tilde{a}$.
Minor deviations are seen for $\DPhi=0$, where the flux does not remain identically zero for increasing $\Wo$.
When $\DPhi=0$, the radial oscillatory flux is generated solely through an inertial mechanism, and increases linearly with $\Wo$ (figure not shown).
The peak radial oscillatory flux generated by this inertial mechanism ($\DPhi=0$) is completely negligible when compared to that generated due to intervillous contractions ($\pi/5 \le \DPhi \le \pi$), by at least an order of magnitude, for all $\Wo$.
The peak oscillatory radial flux therefore remains unaffected by fluid inertia $(\Wo)$ for  $\DPhi > 0$.

\section{Effective velocity boundary condition}
\label{sec:BCeffective}

In this section, we explore an effective velocity boundary condition that captures the primary effects of the pendular-wave along the villi-wall. This boundary condition allows for larger scale simulations of intestinal flow, without explicitly resolving the individual villi. The boundary condition also enables further analytical work in channel flows driven by villi-patterned walls.

Towards this end, we carry out a Fourier decomposition of the instantaneous velocity signal measured at $\bm{u} (t,r \approx H)$ in our simulations, just above the villi tips, identifying the significant modes appearing at relevant length scales.
We therefore propose the following component-wise non-dimensional effective velocity boundary model, $\tilde{\bm{u}}^{e}$, for the system,$\tilde{\bm{u}}^{e}$, for the system,

\begin{align}
\label{eq:model_axialWallVel}
\tilde{u}^{e}_z(t, r\approx H) &=  c_0 + c_1 \sin \left( 2 \pi \left( \frac{t}{T} + \frac{z}{L_z} \right) - \frac{\DPhi}{2} \right) \notag \\
&\qquad - c_2 \sin \left( 2 \pi \frac{z}{P} \right) \cos \left( 2 \pi \left( \frac{t}{T} + \frac{z}{L_z} \right) - \frac{\DPhi}{2} \right)
\end{align}

\begin{align}
\label{eq:model_radialWallVel}
\tilde{u}^{e}_r(t, r\approx H) &= d_0 - d_1 \cos \left( 2 \pi \left( \frac{t}{T} + \frac{z}{L_z} \right) - \frac{\DPhi}{2} \right) \notag \\ 
&\qquad - d_2 \cos \left(  2 \pi \frac{z}{P}  \right) \cos \left( 2 \pi \left( \frac{t}{T} + \frac{z}{L_z} \right) - \frac{\DPhi}{2} \right)
\end{align}

\noindent
where $c_{0-2}$ and $d_{0-2}$, are model coefficients that will depend on $\DPhi$ and $\tilde{a}$, as well on the geometric ratios of the domain.

The effective boundary velocity model consists of three terms for both the axial and radial components.
The first of these is a constant term responsible for irreversible flow, while the remaining two are propagating harmonic terms.
The constant term is set to a non-zero value for the effective axial boundary velocity $\tilde{u}^e_z$ generating a net axial steady streaming flow in the channel while that for the radial $\tilde{u}^e_r$ effective boundary velocity is set to zero.
For both components, the first of the harmonic terms imposes travelling contractions-relaxations at the longer wavelength of the periodic domain $L_z$.
The second harmonic term adds villi scale perturbations, caused by the discrete nature of the villi-wall, to the long-wavelength boundary velocity wave.

The coefficients in (\ref{eq:model_axialWallVel}) and (\ref{eq:model_radialWallVel}) are shown in figure S6 as functions of $\DPhi$ for $\Wo=0.16$.
We obtain the coefficients $c_0$ and $d_0$ by averaging the instantaneous velocity signal measured at $\bm{u}(t, r \approx H)$ from the simulations over both time and space, for different values of $\DPhi$ and $\tilde{a}$.
We see that $c_0$ is non-zero, which is necessary for generating irreversible unidirectional axial flow, manifesting the advected layer.
The measured $c_0$ coefficient is $\approx 90\%$ of $W/(2P)$, the analytically obtained scaling for velocity within the villi zone in (\ref{eq:scalingAxial}).
For the radial component, $d_0 \approx 0$ since the fluid is radially confined by walls and no net radial flow can occur.
The model coefficients can thus be reduced by setting $c_0=\frac{\tilde{a}W}{2P}\sin\left(\DPhi\right)$ and $d_0=0$.
The rest of the fit coefficients ($c_{1,2}$ and $d_{1,2}$) are estimated by least-squares fitting of the model equations to the velocity signal, for all time values, and subsequently taking their averages.
These show a complex dependence on $\DPhi$ and $\tilde{a}$ (see figures S6 and S7).
Since this dependence is non-trivial, we introduce heuristic approximations for these coefficients to achieve reasonable fits for smaller $\DPhi \to 0$.
Note that the variation of $c_2$ and $d_2$ coefficients with $\tilde{a}$, introducing villi scale perturbations to the flow, is consistent with the scaling of the MBL height $\ell$ in section~\ref{sec:MBL:sub:MBLheight}.

\begin{figure}
	\centering
	\includegraphics[width=\textwidth]{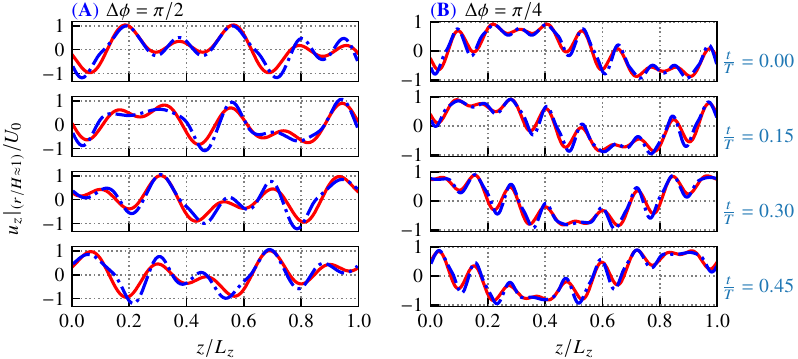}
	\caption{Axial velocity $u_z/U_0$ measured at the villi tips, $r/H \approx 1$, from simulations at $\Wo=0.16$ and $\tilde{a}=0.1$, for \textbf{(A)} $\DPhi=\pi/2$ and \textbf{(B)} $\DPhi=\pi/4$ (dot-dashed, blue) compared with corresponding effective velocity model $\tilde{u}^{e}_z$ from equation~\ref{eq:model_axialWallVel} (solid, red), for four increasing time-fractions. A comparison of the radial velocity $u_r/U_0$ measured at the villus tips with the effective boundary condition is provided in figure S5. The coefficients $c_{0-2}$ for the effective velocity curves plotted here are obtained from the heuristic approximations for the data shown in figure S6.}
	\label{fig:effectiveAxialVelBC}
\end{figure}

We compare the modelled effective axial boundary velocity ($\tilde{{u}}^e_{z}$) with the simulated instantaneous axial velocity ($\tilde{u}_{z}$) at the villi tips ($r=1.02H$) for $\Wo=0.16$ for two $\DPhi$ values, in figure~\ref{fig:effectiveAxialVelBC}.
Similar comparison for the radial component given in (\ref{eq:model_radialWallVel}) is provided in the supplementary figure S5.
Here we have used the functional approximations of the coefficients shown in figure S6 (solid line) to obtain the effective velocity curves in figure~\ref{fig:effectiveAxialVelBC} and figure S5.
The comparisons show excellent agreement with the simulations.
We also note that the model fit improves significantly for the smaller value of $\DPhi=\pi/4$.
We find that the proposed effective velocity boundary condition works well in the Stokes flow regime, but its accuracy decreases for $\Wo \gtrsim 1.0$.

It is worth noting that the effective velocity model in (\ref{eq:model_axialWallVel}) and (\ref{eq:model_radialWallVel}) indicates a  phase-lock between axial and radial velocity oscillations at the scale of $L_z$.
This phase-lock is responsible for the unequal clockwise and counter-clockwise vortical flows within the mixing layer, as seen in figures~\ref{fig:instantVel_lowWo} and \ref{fig:instantVel_highWo}.
Furthermore, one can transform (\ref{eq:model_axialWallVel}) and (\ref{eq:model_radialWallVel}) in the reference frame of the propagating wave, by substituting $z^{R} = z + c t$.
Under such transformation we immediately see that the perturbations occurring at the villi scale (terms associated with $c_2$ and $d_2$) will travel in the $+z$ direction, at the wave speed $c$.
Our proposed model therefore captures the essential physics of villi scale non-reciprocal motion responsible for irreversible flow in the $+z$ direction.

\section{Conclusions}
\label{sec:conclusions}

We investigate the time-periodic flow induced by pendular-wave activity in a channel patterned with villi-like micro-structures, using two-dimensional lattice Boltzmann simulations.

We reveal the existence of oscillatory boundary layer that divides the flow into mixing layer nearer the villi  tips and an advected layer nearer the channel center.
The mixing layer is dominated by asymmetrical semi-vortical flow structures originating at the nodes and anti-nodes of the pendular-wave, and the advected layer shows uniform unidirectional axial flow. The height of this layer scales as $\ell \propto \DPhi^{-2/3}$ in the viscous dominated flow regime, and scales as $\ell \propto \DPhi^{-1/3}$ in the inertia dominated regime, where $\DPhi$ is the phase lag of the villi-wall boundary. We delineate the two flow regimes by achieving a double power law scaling for the measured $\ell$.

We show that the emergence of the advected layer above the mixing layer results from a competition between the decay of the oscillatory axial flow and the unidirectional steady axial flow along the channel.
We map the mixing layer decay with increasing oscillatory inertia for the villi-patterned wall as $\StLayer^{-1/2}$, a departure from the classical scaling of $\StLayer^{-1}$ \citep{schlichting_BoundaryLayerTheory_1960} above oscillating planar walls.
With increased inertia, the axial fluid gets dynamically confined to nearer the villi-tips while the luminal flow velocities increase.
This results in a non-monotonic axial pumping effect, with a maximum seen at $\Wo \approx 2.8$.

We reveal that the advective layer transports fluid in a direction opposite to that of wave propagation; a counter-intuitive phenomenon that persists even at low Womersley numbers.
We demonstrate that this reversal stems from a subtle mechanism rooted in the \textit{scallop theorem} for Stokes flow: the phase lag between adjacent villi imposes non-reciprocal motion on the fluid entrapped in the intervillous spaces, thus breaking the time-reversibility of Stokes equations.
By analogy with the simple 2D Stokes swimmer of \cite{najafi_SimpleSwimmerLow_2004}, we formulate mechanistic scaling laws that capture the direction and magnitude of the irreversible flux through the channel.

Finally, we develop an effective boundary condition at the villus tips that incorporates both the larger villi-array length scale as well as individual villi scale. The effective condition successfully models both radial and axial villi-tip velocities for $\Wo \le 1$, enabling coarse-grained modelling of intestinal flows at the organ scale without the need to explicitly resolve individual villi.

These results shed new light on the flow consequences of both propagating longitudinal contractions and the presence of villi in the small intestine \citep{melville_LongitudinalContractionsDuodenum_1975, lammers_SpatialTemporalCoupling_2005, lentle2012comparison,de2013fluid,fullard_PropagatingLongitudinalContractions_2014}. On the one hand, villi have long been assumed to play a passive role in absorption, primarily by increasing the available surface area \citep{strocchi1993role}. On the other hand, seminal theoretical modelling studies have demonstrated that efficient mixing in the small intestine requires conditions akin to a longitudinal array of perfectly mixed stirred-tank reactors, with slow axial transport of reactants between them \citep{penry1986chemical, penry1987modeling}.
While it is well established that pendular contractions promote local dispersion via shear dispersion mechanisms \citep{de2013fluid,fullard_PropagatingLongitudinalContractions_2014}, our results highlight that villi themselves could play an active role in enhancing transport and mixing. However, our study also reveals that the physiological basis of intestinal motility at the scale of villi remains largely undocumented. In particular, quantitative data describing the amplitude, frequency, and spatio-temporal organization of villus-scale motions are still lacking. Identifying these missing physiological inputs is essential to advance our mechanistic understanding of transport and mixing phenomena in the intestinal lumen.
Our simulations also provides a foundation for future investigations into physiologically relevant effects, such as the influence of non-Newtonian and heterogeneous digesta, as well as radially asymmetric contractile patterns, on transport mechanics in the gastrointestinal tract. 

Future work should extend these 2D simulations to 3D geometries to quantify how villus-scale architecture influences intestinal transport. The transport mechanisms identified here are expected to hold in 3D, as the scaling law is based on inter-villus mass conservation, independent of villus shape or dimensionality. When moving to a cylindrical geometry, we expect from mass conservation arguments that circumferential confinement will enhance the advective layer near the tube center. Among geometric parameters, the lateral spacing between villi—both longitudinal and circumferential—appears particularly critical, controlling the thickness of the steady streaming boundary layer \citep{puthumana2022steady}, and should therefore guide future 3D studies. These extensions will enable a more complete quantitative understanding of mixing and axial transport in realistic intestinal geometries.

Beyond these physiological considerations, these findings suggest that bio-inspired ``pendular-wave'' villus-wall motion could be harnessed for microfluidic applications. Much like artificial cilia arrays \citep{shields2010biomimetic, den2008artificial}, such motion could enable efficient flow control in microfluidic systems. In this context, varying the villi geometry or spacing ($R/H, P/W$) could provide an additional means to tune the fluid fluxes, as seen in related studies \citep{puthumana2022steady, hall_MechanicsCiliumBeating_2020}. This opens new avenues for biomimetic design, where pulsatile boundary and pressure forcing offer an additional degree of freedom for flow control.

\backsection[Supplementary data]{\label{SupMat} Supplementary movies are available.
The simulation code is open source and available at \cite{vernekar20253dintestinalflow}.
}

\backsection[Acknowledgements]{ RV thanks Irina Ginzburg for her advise on LBM advanced boundary condition, as well as J\'er\'emy O'Byrne and Anjishnu Choudhury for helpful discussions on irreversibility in the Stokes flow regime.}

\backsection[Funding]{ Most of the computations presented in this paper were performed using the GRICAD infrastructure (https://gricad.univ-grenoble-alpes.fr), which is supported by Grenoble research communities. LRP is part of the LabEx Tec21 (ANR-11-LABX-0030) and of the PolyNat Carnot Institute (ANR-11-CARN-007-01). The authors thank Agence Nationale de la Recherche for its financial support of the projects TransportGut, ANR-21-CE45-0015 and TABAG, ANR-20-CE30-0001-01.}

\backsection[Declaration of interests]{ The authors report no conflict of interest.}

\backsection[Author ORCIDs]{ R. Vernekar, https://orcid.org/0000-0002-3166-6564;
C. de Loubens, https://orcid.org/0000-0002-4988-9168;
C. Loverdo, https://orcid.org/0000-0002-0888-1717;
Martin Garic, https://orcid.org/0000-0002-5871-3754;
Dácil Idaira Yánez Martín, https://orcid.org/0009-0005-5642-0033;
}

\backsection[Author contributions]{RV and FA performed numerical computations. RV and MG developed theorerical scalings. DIYM and ST provided experimental inputs. RV and CdL carried out analysis of the results. RV authored the computational code.  CL, ST and CdL supervised the research and managed funding. All authors discussed the results and contributed to the final manuscript.}

\appendix

\section{Geometric scaling of radial and axial fluxes}
\label{secX:geometricScalings}
Let the size of any intervillous gap between two adjacent villi along $\hat{\bm{z}}$ be given as,
\begin{equation}
	\label{eqX:gapWithPos}
	\zeta_i = X_{i+1} - X_{i}
\end{equation}
Substituting from (\ref{eq:villiOscPos}) and simplifying we have,
\begin{equation}
	\label{eqX:gapWithPhaseLag}
	\zeta_i = P-W + 2 a \sin \left( \frac{\DPhi}{2} \right) \sin \left ( \omega t + \frac{\DPhi}{2} (2i-1) \right),
\end{equation}
and the expansion (or contraction) rate of the gap is,
\begin{equation}
	\label{eq:gapExpansionRate}
	\dot{\zeta}_i = 2 a \omega \sin \left( \frac{\DPhi}{2} \right) \cos \left ( \omega t + \frac{\DPhi}{2} (2i-1) \right).
\end{equation}

\subsection{Scaling peak oscillatory radial flux}
\label{secX:geometricScalings:sub:radialFluxOsc}
The instantaneous flux driven by the contraction/expansion of any intervillous gap is $|\dot{\zeta}_i|H$.
We postulate that the peak oscillatory radial flux per intervillous gap would scale as,
\begin{equation}
	Q_r^{os,\max} = \frac{1}{2T} \int_t^{t+T} |\dot{\zeta}_i|H dt = \frac{a\omega H}{T} \sin\left(\frac{\DPhi}{2}\right) \int_t^{t+T} \left| \cos \left( \omega t + \frac{\DPhi}{2} (2i-1)\right) \right| dt,
\end{equation}
\noindent
where, we make use of the fact that $\sin(\DPhi/2) > 0$ for $0 \le \DPhi \le \pi$.
Now, the integral can be simplified to,
\begin{equation}
	Q_r^{os,\max} = \frac{2 a\omega H}{T} \sin\left(\frac{\DPhi}{2}\right) \int_{-T/4}^{T/4} | \cos( \omega t ) | dt = \frac{2 a \omega H}{\pi} \sin\left(\frac{\DPhi}{2}\right),
\end{equation}
which gives the non-dimensional oscillatory radial flux as,
\begin{equation}
	\tilde{Q}_r^{os,\max} = \frac{2}{\pi} \sin\left(\frac{\DPhi}{2}\right),
\end{equation}

\subsection{Scaling irreversible axial flux in villi zone}
\label{secX:geometricScalings:sub:AxialSSFluxVilli}
The velocity at the center of the gap bounded by the $\ith$ and $\ith[(i+1)]$ villi using linear interpolation, and further trigonometric simplification is,
\begin{equation}
	\label{eqX:gapFluidVelocity}
	V_i = \frac{U_i+U_{i+1}}{2} = \omega a \cos \left( \frac{\DPhi}{2}\right) \sin \left( \omega t + \frac{\DPhi}{2} (2i-1) \right)
\end{equation}
Now, since the axial flux trough each gap would be proportional to the gap size and fluid velocity within, we compute the velocity weighted by the gap size as,
\begin{align}
	\label{eqX:momentGapSizeVelocity}
	M_{z,i} = \zeta_i V_i &=(P-W) \omega a \sin \left( \omega t + (2i-1) \frac{\DPhi}{2} \right) \cos \left( \frac{\DPhi}{2} \right) \notag \\ &\quad+ a^2 \omega \sin^2 \left( \omega t + (2i-1) \frac{\DPhi}{2} \right) \sin \left( \DPhi \right) .
\end{align}
Taking the time integral for the irreversible component, only the second term in (\ref{eqX:momentGapSizeVelocity}) survives, and can be simplified to,
\begin{equation}
	\label{eqX:SSFMomentSimlpified}
	\overline{M}_z = \frac{1}{T} \int_{t}^{t+T} M_{z,i} dt = \frac{a^2 \omega}{2} \sin \left( \DPhi \right)
\end{equation}
The scaling for the steady streaming irreversible flux through any gap within the villi zone can be now be written as,
\begin{equation}
	Q^{ss,v}_z = \frac{H N}{L_z} \overline{M}_z = \frac{a^2 \omega H}{2 P} \sin \left( \DPhi \right).
\end{equation}
And when non-dimensionalized with characteristic villi flux ($U_0 H = a\omega H$), this is expressed as,
\begin{equation}
	\tilde{Q}^{ss,v}_z = \frac{\tilde{a} W}{2 P} \sin \left( \DPhi \right).
\end{equation}

\bibliographystyle{jfm}
\bibliography{2DVilliFlow}


\end{document}